\documentclass[letterpaper,twocolumn,10pt]{article}
\usepackage{zhanggroup}

\usepackage{tikz}
\usepackage{amsmath}
\usepackage{amsthm}
\usepackage{bbm}
\usepackage{multirow}
\usepackage{graphicx}
\usepackage{booktabs}
\usepackage{adjustbox}
\usepackage{makecell}
\usepackage{xspace}
\usepackage{bbding}
\usepackage{colortbl}
\usepackage{url}

\newcommand{\dalle}{DALL$\cdot$E~2\xspace}
\newcommand{\mypara}[1]{\smallskip\noindent{\bf {#1}.} \xspace}
\begin{document}
%-------------------------------------------------------------------------------

\date{}
\title{\Large \bf DE-FAKE: Detection and Attribution of \\ Fake Images Generated by Text-to-Image Generation Models}

\author{
Zeyang Sha\textsuperscript{1}\ \ \
Zheng Li\textsuperscript{1}\ \ \
Ning Yu\textsuperscript{2}\ \ \
Yang Zhang\textsuperscript{1}
\\
\\
\textsuperscript{1}\textit{CISPA Helmholtz Center for Information Security}\ \ \ 
\textsuperscript{2}\textit{Salesforce Research}
}

\maketitle

%-------------------------------------------------------------------------------
\begin{abstract}
%-------------------------------------------------------------------------------

Text-to-image generation models that generate images based on prompt descriptions have attracted an increasing amount of attention during the past few months.
Despite their encouraging performance, these models raise concerns about the misuse of their generated fake images.
To tackle this problem, we pioneer a systematic study on the detection and attribution of fake images generated by text-to-image generation models.
Concretely, we first build a machine learning classifier to detect the fake images generated by various text-to-image generation models.
We then attribute these fake images to their source models, such that model owners can be held responsible for their models' misuse.
We further investigate how prompts that generate fake images affect detection and attribution.
We conduct extensive experiments on four popular text-to-image generation models, including \dalle, Stable Diffusion, GLIDE, and Latent Diffusion, and two benchmark prompt-image datasets.
Empirical results show that (1) fake images generated by various models can be distinguished from real ones, as there exists a common artifact shared by fake images from different models;
(2) fake images can be effectively attributed to their source models, as different models leave unique fingerprints in their generated images;
(3) prompts with the ``person'' topic or a length between 25 and 75 enable models to generate fake images with higher authenticity.
All findings contribute to the community's insight into the threats caused by text-to-image generation models. 
We appeal to the community's consideration of the counterpart solutions, like ours, against the rapidly-evolving fake image generation.

%-------------------------------------------------------------------------------
\end{abstract}
%-------------------------------------------------------------------------------

%-------------------------------------------------------------------------------
\section{Introduction}
%-------------------------------------------------------------------------------

Text-to-image generation models have made tremendous progress during the past few months.
State-of-the-art models in this field, like Stable Diffusion~\cite{RBLEO22} and \dalle~\cite{RPGGVRCS21}, are able to generate high-quality images ranging from artworks to photorealistic news illustrations. 
Traditional image generation models, such as generative adversarial networks (GANs)~\cite{GPMXWOCB14}, generate synthetic/fake images with latent code sampled from a Gaussian distribution.
On the other hand, text-to-image generation models require users to provide textual inputs, namely \emph{prompts}, and generate images that match the prompts.

The high-quality synthetic images created by text-to-image generation models can be used for various purposes.
For instance, they can facilitate the materialization of a novelist's envisioned scene, perform the automated generation of illustrations for advertising campaigns, and create physical scenes that cannot be captured photographically.
However, these synthetic images also pose severe threats to society.
For instance, such images can be used by malicious parties to disseminate misinformation.
As reported by TechCrunch, Stable Diffusion is able to generate realistic images, e.g., images on the war in Ukraine, that may be used for propaganda.\footnote{\url{https://techcrunch.com/2022/08/12/a-startup-wants-to-democratize-the-tech-behind-dall-e-2-consequences-be-damned/}.}
Also, these images can jeopardize the art industry.
BBC reported that some fake artworks generated by text-to-image generation models won first place in an art competition, which caused the complaints of the involved artists.\footnote{\url{https://www.bbc.com/news/technology-62788725}.}

\begin{figure}[!t]
\centering
\includegraphics[width=\columnwidth]{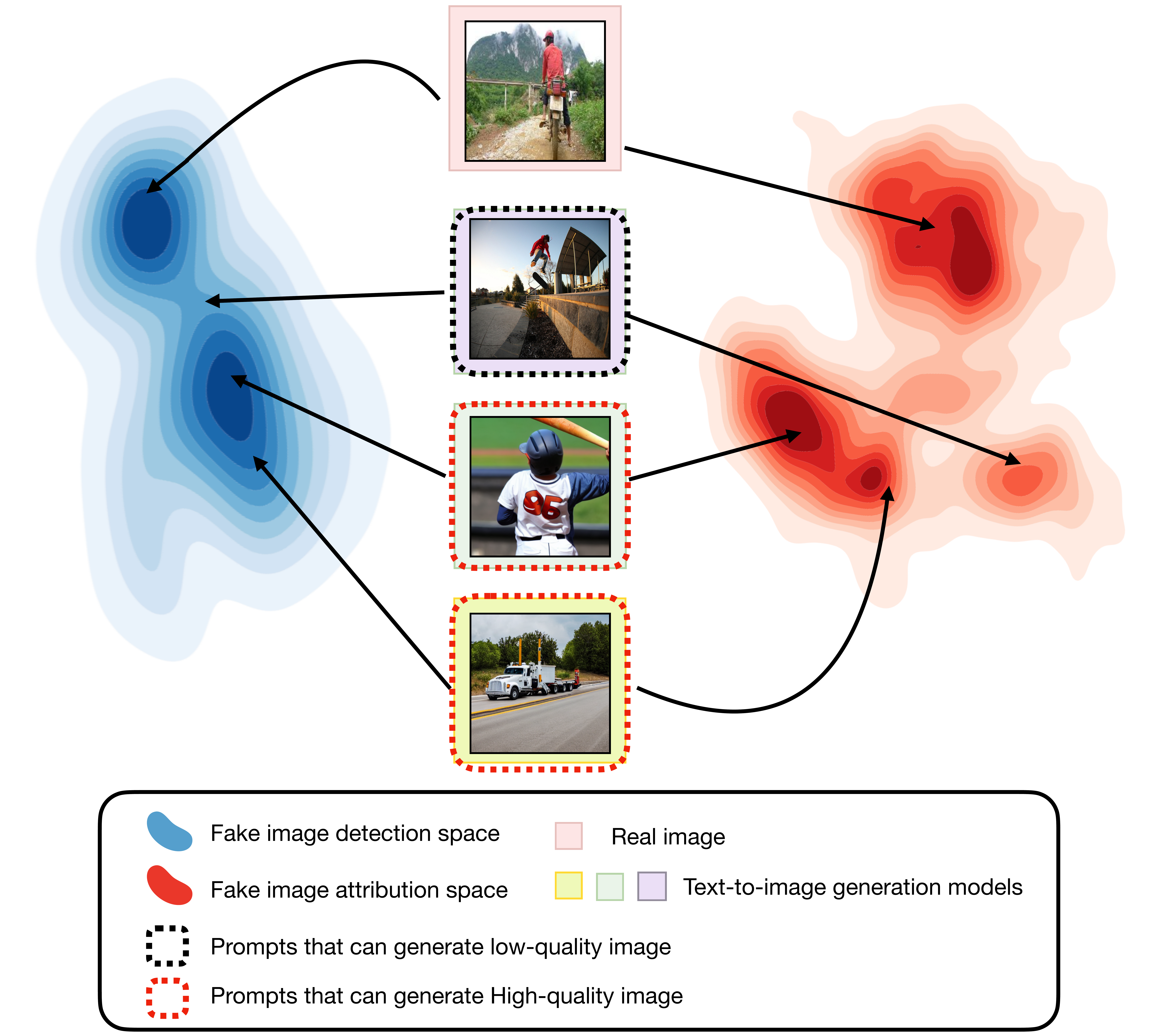}
\caption{An illustration of our work, including fake image detection, fake image attribution, and prompt analysis.}
\label{fig:overview}
\end{figure}

%-------------------------------------------------------------------------------
\subsection{Our Contributions}
%-------------------------------------------------------------------------------

There are multiple approaches to alleviate the concerns brought by advanced generation models.
In particular, one can build a detector to detect whether an image is real or fake automatically.
Moreover, one can build an attribution model to attribute a synthetic image to its source generation model, such that the model owner can be held responsible for the model's misuse.
So far, various efforts have been made in this field; however, they only focus on traditional generation models, represented by GANs~\cite{WWZOE20, YDF19, GSRS21}.
To the best of our knowledge, no study has been done on text-to-image generation models.
Also, whether the prompts used in such models can facilitate fake image detection and attribution remains unexplored.

In this work, we present the first study on the detection and attribution of fake images generated by text-to-image generation models.
Concretely, we formulate the following three research questions (\textbf{RQ}s).
\begin{itemize}
\item \mypara{RQ1}
Can we differentiate the fake images generated by various text-to-image generation models from the real ones, i.e., detection of fake and real images?
\item \mypara{RQ2}
Can we attribute the fake images to their source text-to-image generation models, i.e., attribution of fake images to their sources?
\item \mypara{RQ3}
What kinds of prompts are more likely to generate authentic images?
\end{itemize}

\mypara{Methodology} 
To differentiate fake images from real ones (\textbf{RQ1}), i.e., fake image detection, we train a binary classifier/detector.
To validate the generalizability of the detector, we especially train it on fake images generated by \textit{only one} model and evaluate it on fake images generated by \textit{many other} models.
We consider two detection  methods, i.e., \emph{image-only} and \emph{hybrid}, depending on the detector's knowledge.
The image-only detector makes its decision solely based on the image itself.
The hybrid detector considers both images and their corresponding prompts.
Hybrid detection is a brand-new detection method, and it is designed specifically for detecting fake images created by text-to-image generation models.
Concretely, we leverage the image and text encoders of the CLIP model~\cite{RKHRGASAMCKS21} to transfer an image and its prompt to two embeddings which are then concatenated as the input to the detector.
Note that in the prediction phase, an image's natural prompt may not be available.
In such cases, we leverage an image captioning model BLIP~\cite{LLXH22} to generate the prompt for the image.

To attribute a fake image to its source model (\textbf{RQ2}), we propose \textit{fake image attribution} by training a multi-class classifier (instead of a binary classifier), and we name this classifier as an attributor.
Specifically, the attributor is trained on fake images generated by multiple text-to-image generation models.
Fake images from the same model are labeled as the same class.
Moreover, we also establish the attributor by two methods, i.e., image-only and hybrid, which are the same as the detector to address \textbf{RQ1}.

Different from \textbf{RQ1} and \textbf{RQ2}, \textbf{RQ3} focuses on the impact of prompts on the authenticity of generated images.
To this end, we conduct \textit{prompt analysis} from semantic and structural perspectives.
In the former, we design two semantic extraction methods to analyze the impact of prompt topics on the authenticity of fake images.
More specifically, the first one directly uses the ground truth topics provided in the dataset for each prompt, and the second one automatically clusters the various prompts into different groups and extracts topics from these groups.
From the structural perspective, we conduct the study based on the length of prompts and the proportion of nouns in prompts, respectively.
\autoref{fig:overview} presents an overview of our methods to address the three research questions.

\mypara{Evaluation}
We perform experiments on two benchmark prompt-image datasets including MSCOCO~\cite{LMBHPRDZ14} and Flickr30k~\cite{YLHH14}, and four popular pre-trained text-to-image generation models including Stable Diffusion~\cite{RBLEO22}, Latent Diffusion~\cite{RBLEO22}, GLIDE~\cite{NDRSMMSC21}, and \dalle~\cite{RDNCC22}.

In fake image detection, extensive experimental results show that image-only detectors can achieve good performance in some cases, while hybrid detectors can always achieve better performance in all cases.
For example, on MSCOCO~\cite{LMBHPRDZ14}, the image-only detector trained on fake images generated by Stable Diffusion can achieve an accuracy of 0.834 in differentiating fake images generated by Latent Diffusion from the real ones, while it can only achieve 0.613 and 0.554 on GLIDE and \dalle, respectively.
Encouragingly, the hybrid detector trained on fake images from Stable Diffusion achieves 0.932/0.899/0.885 accuracy on Latent Diffusion/GLIDE/\dalle with natural prompts, and 0.945/0.909/0.891 accuracy with BLIP-generated prompts.
These results demonstrate that fake images from various models can indeed be distinguished from real images.
We further extract a common feature from fake images generated by various models in \autoref{section:detection-disscussion}, which implies the existence of a common artifact shared by fake images across various models.

In fake image attribution, our experiments show that both image-only and hybrid attributors can achieve good performance in all cases. 
Similarly, the hybrid attributor is better than the image-only one.
For instance, the image-only attributor can achieve an accuracy of 0.815 in attributing fake images to the models we consider, while the hybrid attributor can achieve 0.880 with natural prompts and 0.850 with BLIP-generated prompts.
These results demonstrate that fake images can indeed be attributed to their corresponding text-to-image generation models.
We further show the unique feature extracted from each model in \autoref{section:source-discussion}, which implies that different models leave unique fingerprints in the fake images they generate.

In prompt analysis, we first find that prompts with the topics of ``skis,'' and ``snowboard'' tend to generate more authentic images through our first semantic extraction method, which relies on the ground truth information from the dataset.
However, by clustering various prompts over embeddings by sentence transformer~\cite{RG19}, we find that prompts with the ``person'' topic can actually generate more authentic images.
Upon further inspection, we discover that most of the images associated with ``skis'' and ``snowboard'' are also related to ``person.''
These results indicate that prompts with the topic ``person'' are more likely to generate authentic fake images.
From the structural perspective, our experiments show that prompts with a length between 25 and 75 enable text-to-image generation models to generate fake images with higher authenticity, while the proportion of nouns in the prompt has no significant impact.

\mypara{Implications}
In this paper, we make the first attempt to tackle the threat caused by fake images generated by text-to-image generation models.
Our results on detecting fake images and attributing them to their source models are encouraging.
This suggests that our solution has the potential to play an essential role in mitigating the threats.
We will share our source code with the community to facilitate research in this field in the future.

%-------------------------------------------------------------------------------
\section{Preliminaries}
%-------------------------------------------------------------------------------

%-------------------------------------------------------------------------------
\subsection{Text-to-Image Generation Models}
%-------------------------------------------------------------------------------

During the past few months, text-to-image generation models have attracted an increasing amount of attention. 
A model in this domain normally takes a prompt, i.e., a piece of text, and a random noise as the input and then denoises the image under the guidance of the prompt so that the generated image matches the description.
In this work, we focus on four popular text-to-image generation models that are publicly available online.

\begin{itemize}
\item \mypara{Stable Diffusion~\cite{RBLEO22}} 
Stable Diffusion (SD) is a diffusion model for text-to-image generation.
The available model we use\footnote{\url{https://github.com/CompVis/stable-diffusion}.} is pre-trained on 512$\times$512 images from a subset of the LAION-5B~\cite{SVBKMKCJK21} dataset.
The CLIP model's~\cite{RKHRGASAMCKS21} text encoder is used to condition the model on prompts.
\item \mypara{Latent Diffusion~\cite{RBLEO22}} 
Latent Diffusion (LD) is also a diffusion model for text-to-image generation.
The available model we use\footnote{\url{https://github.com/CompVis/latent-diffusion}.} is pre-trained on LAION-400M, a smaller dataset sampled from LAION-5B~\cite{SVBKMKCJK21}.
Latent Diffusion also leverages the text encoder of CLIP to guide the direction of fake images.
\item \mypara{GLIDE~\cite{NDRSMMSC21}} 
GLIDE is a text-to-image generation model proposed by OpenAI.
The available model\footnote{\url{https://github.com/openai/glide-text2im}.} is trained on a filtered version of a dataset comprised of several hundred million prompt-image pairs.
In addition, GLIDE is not good at understanding prompts that contain ``person'' topics, as such images have been removed from the training dataset due to ethical concerns.
\item \mypara{\dalle~\cite{RDNCC22}} 
\dalle is one of the most popular text-to-image generation models proposed by OpenAI.
A transformer~\cite{VSPUJGKP17} is used to capture the information from both images and prompts.
In our experiments, we adopt a Pytorch version of \dalle released by \dalle-Pytorch.\footnote{\url{https://github.com/lucidrains/DALLE2-pytorch}.}
\dalle-Pytorch uses an extra layer of indirection with the prior network and is trained on a subset of LAION-5B~\cite{SVBKMKCJK21}.
Note that we refer to \dalle-Pytorch as \dalle in the rest of our paper.
\end{itemize}

%-------------------------------------------------------------------------------
\subsection{Datasets}
%-------------------------------------------------------------------------------

We use the following two benchmark prompt-image datasets to conduct our experiments.
\begin{itemize}
\item \mypara{MSCOCO~\cite{LMBHPRDZ14}}
MSCOCO is a large-scale objective, segmentation, and captioning dataset.
It is a standard benchmark dataset for evaluating the performance of computer vision models.
MSCOCO contains 328,000 images distributed in 80 classes of natural objects, and each image in MSCOCO has several corresponding captions, i.e., prompts.
In this work, we consider the first 60,000 prompts due to the constraints of our lab's computational resources.
\item \mypara{Flickr30k~\cite{YLHH14}}
Flickr30k is a widely used dataset for research on image captioning, language understanding, and multimodal learning. 
It contains 31,783 images and 158,915 English prompts on various scenarios.
All images are collected from the Flickr website, and the prompts are written by Flickr users in natural language.
\end{itemize}
Note that the prompts from these datasets are also important as they will be used to generate fake images or serve as inputs for the hybrid classifiers (see \autoref{section:universal_detection} for more details).

In summary, the text-to-image generation models and datasets we consider in this work are listed in \autoref{table:fake_image_setting}.
Since different models are trained on images of different sizes and fake images usually appear in the real world at different resolutions.
Therefore, we adopt the default settings of these available models and perform experiments on fake images of different sizes.

%-------------------------------------------------------------------------------
\section{Fake Image Detection}
\label{section:universal_detection}
%-------------------------------------------------------------------------------

In this section, we present our fake image detector to differentiate fake images from real ones (\textbf{RQ1}).
We start by introducing our design goals.
Then, we present how to construct the detector.
Finally, we show the experimental results.

\begin{table}[!t]
\centering
\caption{The text-to-image generation models, datasets, and the number/size of fake images we consider in this work.
Note that the number of fake images from \dalle being low is due to its poor image generation efficiency.}
\label{table:fake_image_setting}
\begin{tabular}{l| c c c}
\toprule
\textbf{Model} & \textbf{Dataset} & \textbf{Images} &\textbf{Image Size} \\
\midrule
\multirow{2}*{SD} & MSCOCO & 59,247 & 512$\times$512 \\ 
~ & Flickr30k & 13,231  & 512$\times$512 \\
\midrule
\multirow{2}*{LD} & MSCOCO & 31,276  & 256$\times$256 \\ 
~ & Flickr30k & 17,969  & 256$\times$256 \\
\midrule
\multirow{2}*{GLIDE} & MSCOCO & 41,685  & 256$\times$256 \\ 
~ & Flickr30k & 27,210  & 256$\times$256 \\
\midrule
\multirow{2}*{\dalle} & MSCOCO & 1,028  & 256$\times$256 \\ 
~ & Flickr30k & 300  & 256$\times$256 \\
\bottomrule
\end{tabular}
\end{table}

%-------------------------------------------------------------------------------
\subsection{Design Goals}
%-------------------------------------------------------------------------------

To tackle the threats posed by the misuse of various text-to-image generation models, the design of our detector should follow the following points.

\begin{itemize}
\item \mypara{Differentiating Between Fake and Real Images}
The primary goal of the detector is to effectively differentiate fake images generated by text-to-image generation models from real ones.
Successful detection of fake images can reduce the threat posed by the misuse of these advanced models.
\item \mypara{Agnostic to Models and Datasets}
As text-to-image generation models have undergone rapid development, it is likely that more advanced models will be proposed in the future.
As a result, it is difficult for us to collect all text-to-image generation models to build our detector.
Moreover, building the detector on various models (even though we can collect many) inevitably leads to more resource consumption.
Therefore, it is crucial to explore whether our detection based on very few text-to-image generation models is generalizable to other models.
Also, since we have no knowledge of the distribution of prompts used to generate fake images, it is also important for our detector to identify fake images generated by prompts from other prompt-image datasets.
\end{itemize}

%-------------------------------------------------------------------------------
\subsection{Methodology}
%-------------------------------------------------------------------------------

To achieve the primary goal of differentiating fake images from real ones, we construct a detector by training a binary classifier.
Furthermore, to make our detector agnostic to unseen models and datasets, we consider a more realistic and challenging scenario where the detector can collect fake images generated by only one text-to-image generation model given prompts from one dataset.
The detector then trains its binary classifier on these fake/real images and evaluates its generalizability on fake images from other models and datasets.

\begin{figure}
\centering
\includegraphics[width=\columnwidth]{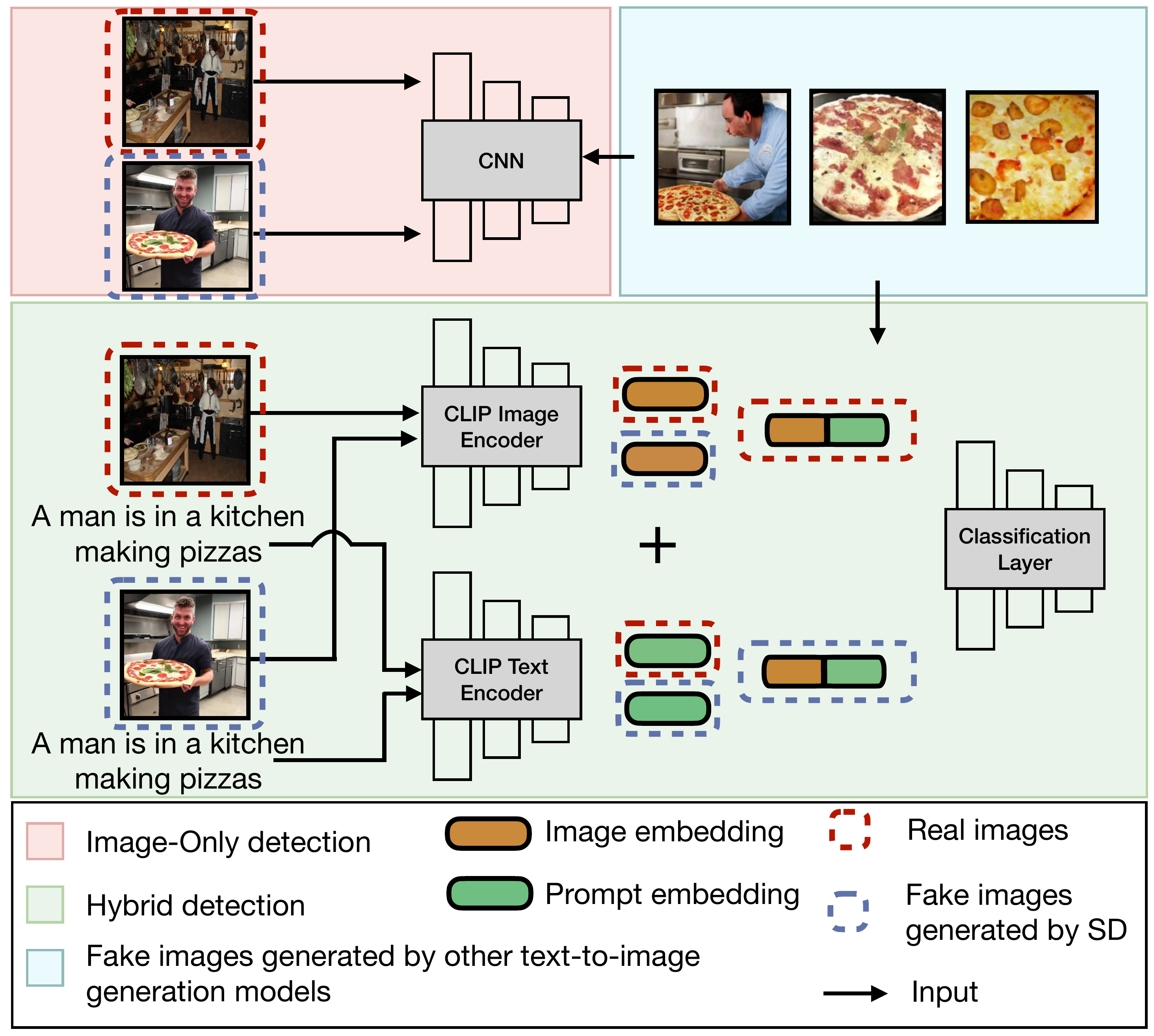}
\caption{An illustration of fake image detection.
The red part describes image-only detection.
The green part describes hybrid detection.
The blue part describes fake images generated by other text-to-image generation models.}
\label{fig:uni_detect}
\end{figure}

In addition, based on the background knowledge available to the detector, we propose two different approaches to establish the detector, namely image-only and hybrid.
The image-only detector accepts only images as input.
In contrast, the hybrid detector accepts both images and their corresponding prompts as input.
See \autoref{fig:uni_detect} for an illustration of how to conduct fake image detection.

\mypara{Image-Only Detection}
The red part of \autoref{fig:uni_detect} shows the work pipeline of our image-only detector.
The process of training our image-only detector can be divided into three stages, namely, data collection, dataset construction, and detector construction.

\begin{itemize}
\item \mypara{Data Collection} 
We first randomly sample 20,000 images from MSCOCO and treat them as real images for the next stage.
Then, we use the prompts of these 20,000 images to query one model (we choose SD here) to get 20,000 fake images.
In this way, our fake images are from one text-to-image generation model given prompts from one dataset, referred to as SD+MSCOCO.
\item \mypara{Dataset Construction} 
We label all fake images as 0 and all real images as 1.
We then create a balanced training dataset containing a total of 40,000 images.
\item \mypara{Detector Construction} 
We build the detector (i.e., a binary classifier) that accepts an image as input and outputs a binary prediction, i.e., 0-fake or 1-real.
Lastly, we train the detector from scratch with fake and real images in conjunction with classical training techniques. 
Note that we use ResNet18~\cite{HZRS16} as our image-only detector's architecture.
\end{itemize}

After we have trained the detector, we evaluate the generalizability of the trained detector on images from other models, i.e., LD, GLIDE, and \dalle, given prompts from the other dataset, i.e., Flickr30k.
For completeness, we also include the detection results on fake images from the same model and/or the same dataset.
\autoref{table:fake_image_setting} shows the total number of fake images generated by four models and two datasets.
Besides the 20,000 images (out of 59,247) from SD+MSCOCO, which are used to train the detector, all the others are used to test the performance of the detector.
Note that in all cases, we sample the same number of real images as the fake ones for training and testing the detector (and the attributor in \autoref{section:source}).

\mypara{Hybrid Detection}
We now present the hybrid detector, which considers both images and their corresponding prompts.
This is motivated by the observation that real images always carry a wide range of contents that the prompts cannot fully and faithfully describe.
However, since fake images are generated based on prompts, they may not contain additional content beyond what is described, i.e., not as informative as real images.
Therefore, introducing prompts together with images enlarges the disparity between fake and real images, which in our opinion can contribute to differentiating between the two.
We further show in \autoref{section:detection-disscussion} that the disparity between real and fake images is indeed huge from the prompt's perspective.
Note that using prompts as an extra signal for fake image detection is novel and unique to text-to-image generation models, as prompts do not participate in the image generation process of traditional generation models, like GANs.

The green part of \autoref{fig:uni_detect} shows the work pipeline of our hybrid detection.
Specifically, the process of training our hybrid detector can also be divided into three stages, i.e., data collection, dataset construction, and detector construction.
\begin{itemize}
\item \mypara{Data Collection}
To collect the real and fake images, we follow the same step as the first step for the image-only detector.
\item \mypara{Dataset Construction}
Since our hybrid detector takes images and prompts as input, we label all fake images and their corresponding prompts as 0 and label real images and their corresponding prompts as 1.
Similarly, we then create a training dataset containing a total of 40,000 prompt-image pairs.
\item \mypara{Detector Construction}
To exploit the prompt information, we take advantage of CLIP's image encoder and text encoder as feature extractors to obtain high-level embeddings of images and prompts.
Then, we concatenate image embeddings and text embeddings together as new embeddings and use these embeddings to train a binary classifier, i.e., a 2-layer multilayer perceptron, as our detector. 
\end{itemize}

To evaluate the trained hybrid detector, we need both images and their corresponding prompts.
Typically, a user may attach a description to an image they post on the Internet.
Therefore, we can directly consider this attached description as the prompt for the image.
In our experiments, we adopt the original/natural prompts from the dataset to conduct the evaluation.

In a more realistic and challenging scenario where the detector cannot obtain the natural prompts, we propose a simple yet effective method to generate the prompts ourselves.
Concretely, we leverage the BLIP~\cite{LLXH22} model (an image captioning model) to generate captions for the queried images and then regard these generated captions as the prompts for the images.

%-------------------------------------------------------------------------------
\subsection{Results}
\label{section;detection-results}
%-------------------------------------------------------------------------------

We now present the performance of our proposed image-only detection and hybrid detection for fake image detection.

\mypara{Image-Only Detection}
For a convincing evaluation, we adopt the existing work~\cite{WWZOE20} on detecting fake images generated by various types of generation models, including GANs and low-level vision models~\cite{CCXK18, DCZXZ19}, as a baseline.
The authors of~\cite{WWZOE20} name their classifier as \textit{forensic classifier}.
Note that this forensic classifier is the state-of-the-art fake image detector for generation models, and the authors show that it has strong generalizability.
For instance, it can achieve an accuracy of 0.946 on differentiating fake images generated by StarGAN~\cite{CCKHKC18}, which is not considered during the model training, from real images.

\begin{figure}[!t]
\centering
\begin{subfigure}{0.49\columnwidth}
\includegraphics[width=\columnwidth]{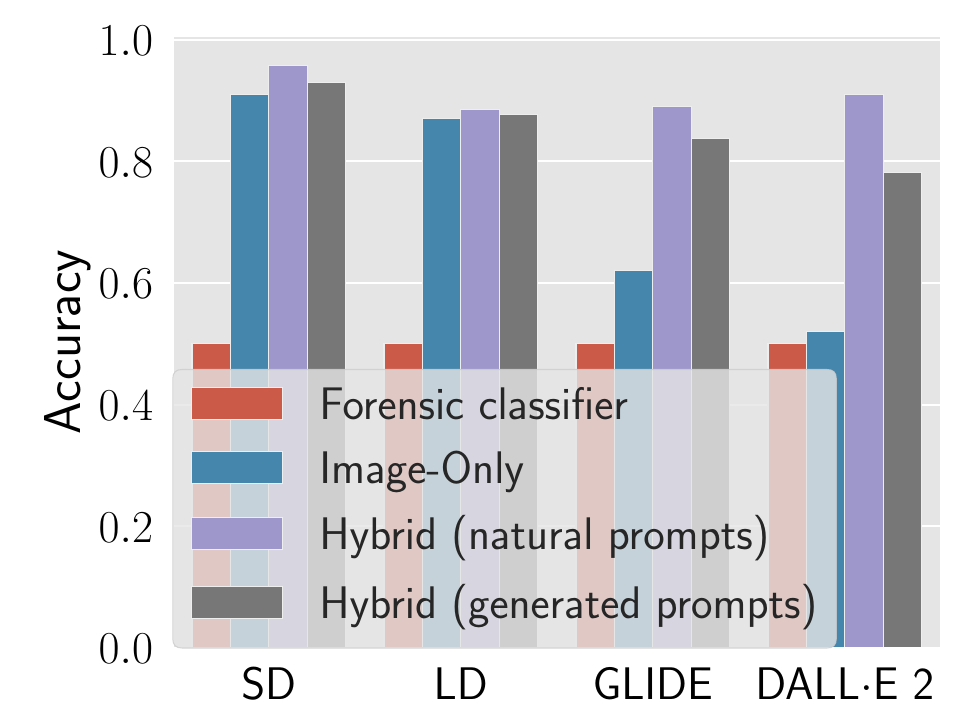}
\caption{MSCOCO}
\label{fig:text_coco}
\end{subfigure}
\begin{subfigure}{0.49\columnwidth}
\includegraphics[width=\columnwidth]{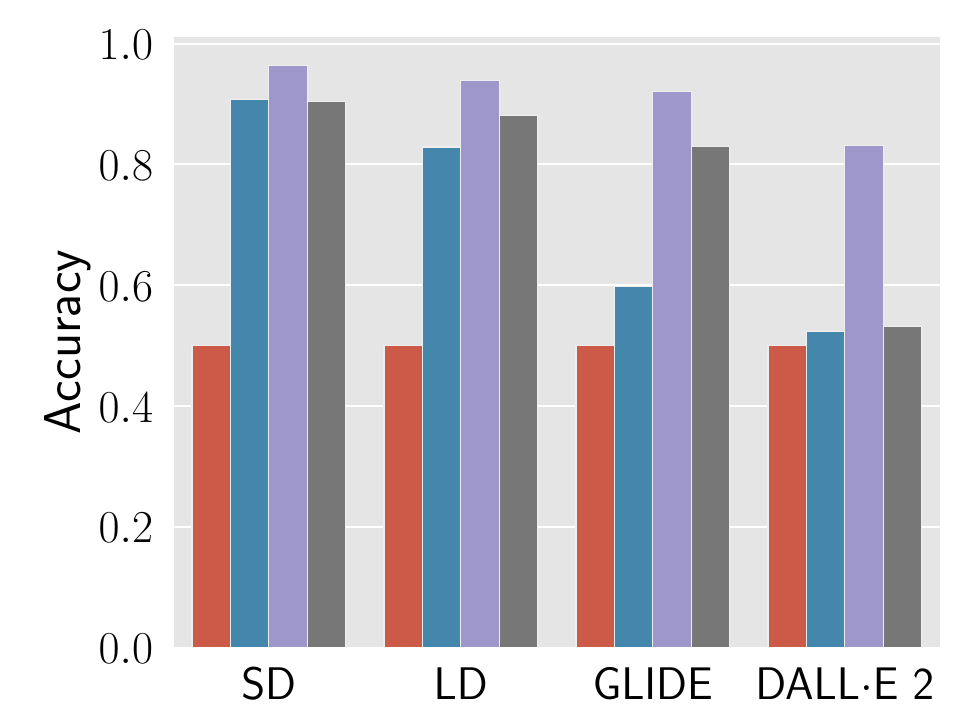}
\caption{Flickr30k}
\label{fig:text_fli}
\end{subfigure}
\caption{The performance of the forensic classifier and detectors.
We conduct the evaluation on (a) MSCOCO  and (b) Flickr30k, respectively.}
\label{fig:text}
\end{figure}

\autoref{fig:text} depicts the evaluation results.
First of all, we can observe that the forensic classifier cannot effectively distinguish fake images (generated by text-to-image generation models) from real ones.
In all cases, the forensic classifier only achieves an accuracy of 0.5, which is equivalent to a random guess.
Based on this observation, we can conclude that the forensic classifier cannot be generalized to text-to-image generation models.
We attribute this observation to the differences between traditional generation models and text-to-image generation models.
This result also prompts the urgent need for counterpart solutions against the misuse of text-to-image generation models.

Furthermore, we can observe that the image-only detector performs much better in all cases than the forensic classifier.
For example, the image-only detector can achieve an accuracy of 0.871 in distinguishing fake images generated by LD+Flickr30k (querying the prompts of Flickr30k to LD) from real images.
We emphasize here that the image-only detector is trained only on fake images generated by SD+MSCOCO and has never seen fake images generated by other models given prompts from other datasets.
We conjecture that this is due to some common properties shared by all fake images generated by text-to-image generation models (see \autoref{section:detection-disscussion} for more details).

Lastly, another interesting finding is the much larger variation in detection performance due to the effect of the model compared to the effect of the dataset.
E.g., in \autoref{fig:text_coco}, the image-only detector achieves an accuracy of 0.913 on SD but only 0.526 on \dalle.
In contrast, comparing \autoref{fig:text_coco} and \autoref{fig:text_fli}, the image-only detector achieves very close accuracy on different datasets over all text-to-image generation models.
We attribute this observation to the unique fingerprint of fake images generated by text-to-image generation models (see \autoref{section:source-discussion}).

\mypara{Hybrid Detection}
Although the image-only detector achieves better performance in all cases compared to the forensic classifier, we acknowledge that the current detection performance is far from the design goal due to the lack of good performance on other models, such as GLIDE and \dalle.
As mentioned earlier, using prompts as an extra signal may boost the fake image detection performance.

We report the performance of our proposed hybrid detection in \autoref{fig:text}.
First, we can find that the hybrid detector can always achieve much better performance than the image-only detector, especially on models like GLIDE and \dalle.
For instance, the hybrid detector with natural prompts can achieve an accuracy of 0.909 on \dalle\!\!+MSCOCO, which is much higher than the 0.522 achieved by the image-only detector.
Moreover, even without natural prompts, the hybrid detector with BLIP-generated prompts can still have a strong performance.
For example, on fake images generated by GLIDE+MSCOCO, the hybrid detector with natural prompts achieves an accuracy of 0.891, and encouragingly, the hybrid detector with BLIP-generated prompts also achieves a high accuracy of 0.838.
These results indicate that introducing prompts together with images can indeed enlarge the disparity between fake and real images, which is beneficial to fake image detection.
We further investigate in more depth why using the prompt as a new signal can improve detection performance (see \autoref{section:detection-disscussion} for detailed information).

Besides, we can find that the performance of the hybrid detector on other models is much less influenced by prior knowledge of the known model than the image-only detector.
For example, on the MSCOCO dataset, the hybrid detector with natural prompts can achieve an accuracy of 0.958 on SD, while the accuracy only drops to 0.909 on \dalle.
We can also find that the hybrid detector is not influenced much by the dataset, similar to the image-only detector.
For instance, on SD, the hybrid detector with generated prompts can achieve quite a similar accuracy between MSCOCO and FLickr30k (0.930 vs.\ 0.904).
These results show that our proposed hybrid detector is strong regarding model and dataset independence.

%-------------------------------------------------------------------------------
\subsection{Discussion}
\label{section:detection-disscussion}
%-------------------------------------------------------------------------------

\begin{figure}[!t]
\centering
\begin{subfigure}{0.47\columnwidth}
\includegraphics[width=\columnwidth]{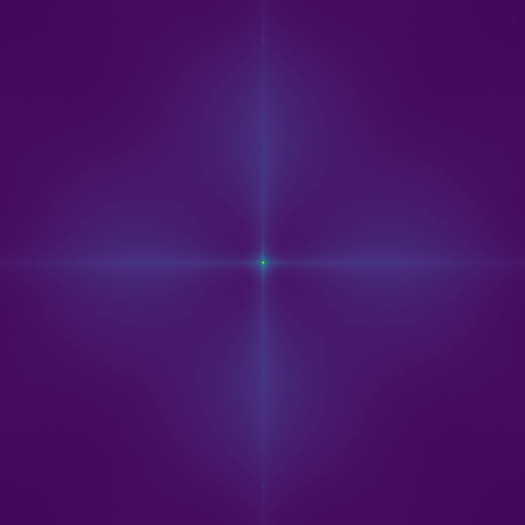}
\caption{Real}
\label{fig:fig_real}
\end{subfigure}
\begin{subfigure}{0.47\columnwidth}
\includegraphics[width=\columnwidth]{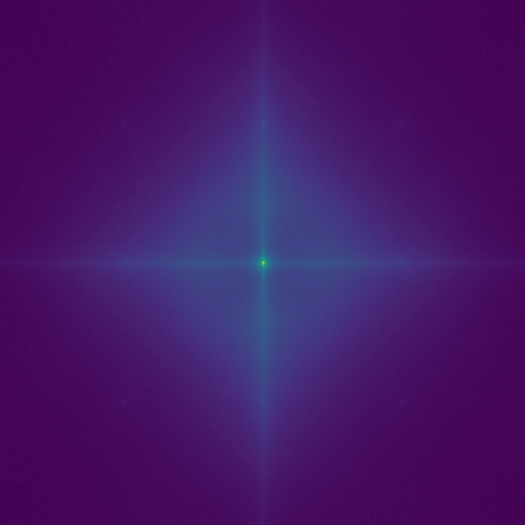}
\caption{Fake}
\label{fig:fig_fake}
\end{subfigure}
\caption{The visualization of frequency analysis on (a) real images and (b) fake images.}
\label{fig:real-fake}
\end{figure}

The above results fully demonstrate the effectiveness of our fake image detection.
Next, we delve more deeply into the reasons for successfully distinguishing fake images from real ones.
We conjecture that there exist some common properties shared by fake images from various text-to-image generation models.
We verify this conjecture by visualizing the common artifact shared across fake images.
Besides, based on the better performance achieved by hybrid detection, we further explore why additional prompt information can enhance detection performance.
In the end, we also test whether our trained detector can be directly applied to fake images from other domains, in particular, fake artwork detection.

\begin{figure}[!t]
\centering
\includegraphics[width=0.8\columnwidth]{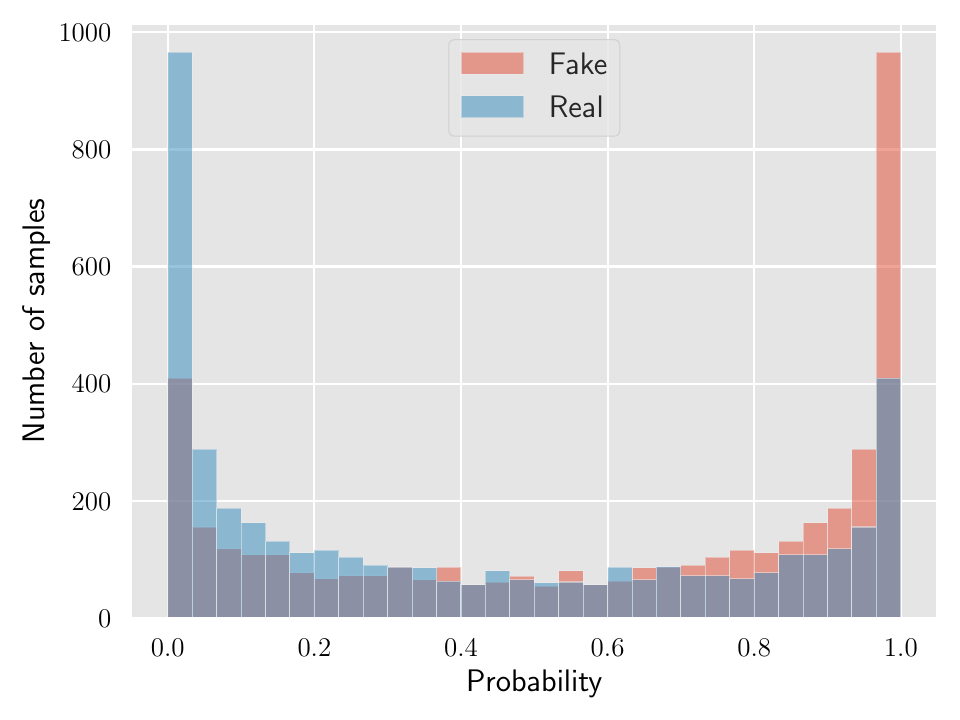}
\caption{The probability distribution of the connection between the real/fake images and the corresponding prompts.}
\label{fig:distance}
\end{figure}

\mypara{Artifact Visualization}
Inspired by Zhang et al.~~\cite{ZKC192}, we draw the frequency spectra of fake and real images.
For the four text-to-image generation models we consider in this work, we randomly select 1,000 fake images from each model given prompts from MSCOCO.
In total, we have obtained 4,000 fake images.
Also, we collect 4,000 real images of the same prompts from MSCOCO.
We then calculate the average of Fourier transform outputs of real and fake images, respectively.
We leverage Fourier transform here due to its ability to reveal latent features of the given images.

As shown in \autoref{fig:real-fake}, we can clearly observe that there are distinct patterns in real and fake images.
Concretely, the central region of the fake image has higher brightness and more concentrated frequency spectra.
This observation verifies the existence of the common artifact shared by the fake images generated by various text-to-image generation models.

\mypara{Why Does Prompt Enhance Detection Performance}
We conduct a more in-depth study on why using prompts as a new signal can improve detection performance.
As mentioned before, a prompt cannot completely reflect the contents of a real image. 
Meanwhile, a fake image is purely based on the prompt information.
This suggests the connection between a fake image and its prompt is stronger than the connection between a real image and its prompt.
This is essentially the reason why the hybrid detector has a better performance than the image-only detector.

To verify this, we first randomly sample 2,000 prompts from MSCOCO.
For each prompt, we collect its corresponding real image from the dataset and let SD generate a fake image for it.
Then, we rely on CLIP's text encoder to transfer the prompt to an embedding and CLIP's image encoder to transfer the real and fake images to two embeddings, respectively.
Then, we calculate two cosine similarities, one is between the prompt's embedding and its real image's embedding, and the other is between the prompt's embedding and its fake image's embedding. 
Finally, the two cosine similarities are normalized into a probability distribution via a softmax function~\cite{RKHRGASAMCKS21}. 
Higher probability implies a stronger connection between the image and the prompt.
\autoref{fig:distance} shows the similarity distribution between the 2,000 prompts and the real/fake images.
We can see that the similarity between the fake image and the corresponding prompt is closer than that between the real image and the same prompt, leading to a clear gap in the similarity distribution between fake and real images.
This verifies our aforementioned intuition.
Furthermore, we can also conclude that it is not the prompt information itself that enhances the performance of the detector, but the prompt information can be exploited as an extra ``anchor'' to provide a new signal to distinguish between real and fake images.
Such signals can be effectively captured by a multilayer perceptron.

\mypara{Case Study of Artwork}
So far, all the previous fake images we have studied are related to MSCOCO and Flickr30k, which are about natural objects.
The experimental results show that our proposed fake image detection can achieve excellent performance in differentiating these fake images from the real ones.
However, text-to-image generation models can also be used to generate other types of images, especially fake artwork.
Therefore, it is interesting to see whether our proposed fake image detection can be directly used to distinguish between real and fake artworks.

Since there do not exist many datasets on artworks, we collect 50 real artworks and 50 fake artworks generated by SD from the Internet.
Besides, since there are no corresponding prompts for these collected artworks, we adopt the image-only detector and the hybrid detector with generated prompts by BLIP to conduct the evaluation.
Note that both detectors we adopt have been trained in previous experiments based on SD+MSCOCO.
The experiments show that our proposed detectors can still achieve good performance in differentiating fake artworks from real ones.
For instance, the image-only detector achieves an accuracy of 0.710, and the hybrid detector can achieve an accuracy of 0.690.
The results, again, indicate that fake images of different styles (e.g., artworks and natural objects) generated by text-to-image generation models share common properties.

%-------------------------------------------------------------------------------
\subsection{Ablation Study}
\label{section:universal_ablation}
%-------------------------------------------------------------------------------

\begin{figure}[!t]
\centering
\begin{subfigure}{0.49\columnwidth}
\includegraphics[width=\columnwidth]{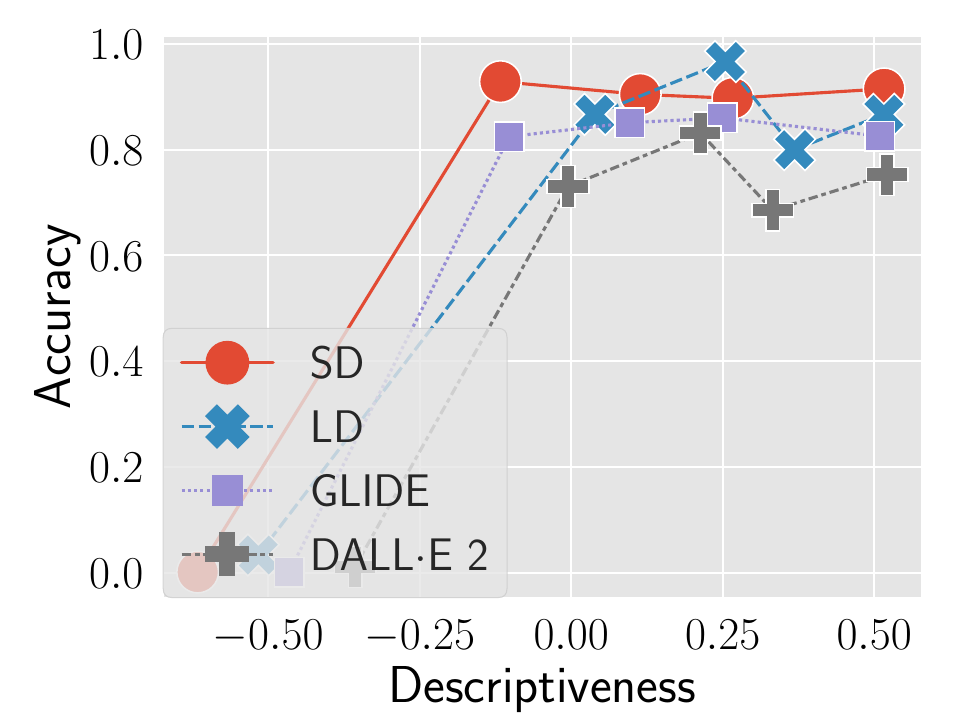} 
\caption{MSCOCO}
\label{fig:des_mscoco}
\end{subfigure}
\begin{subfigure}{0.49\columnwidth}
\includegraphics[width=\columnwidth]{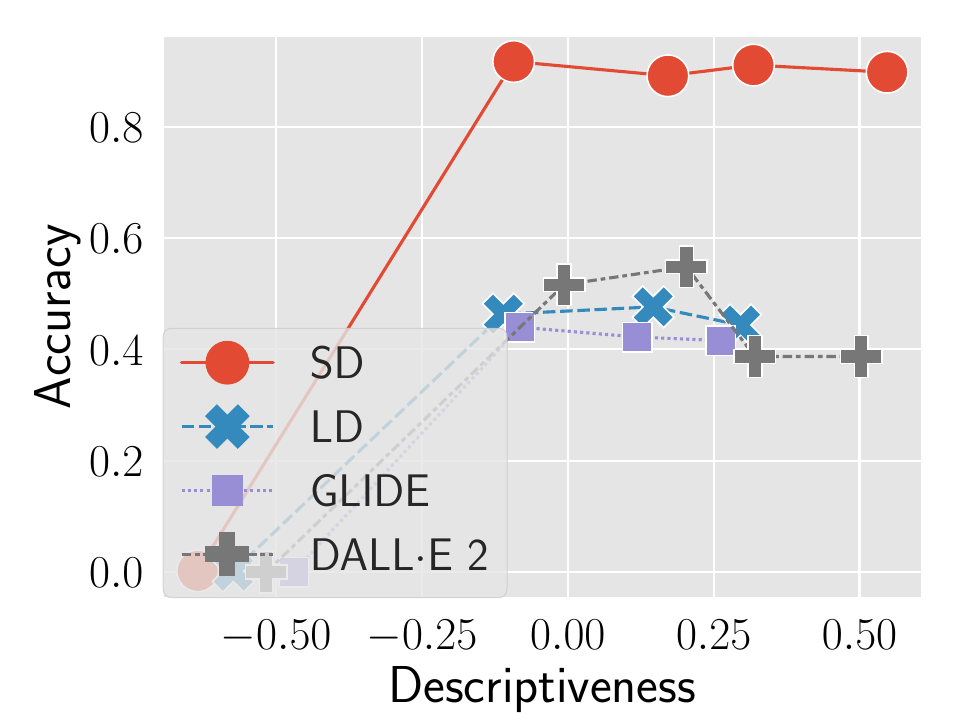}
\caption{Flickr30k}
\label{fig:des_flickr}
\end{subfigure}
\caption{The performance of hybrid detectors with generated prompts in terms of the prompts' descriptiveness.
The descriptiveness is grouped into five equally sized bins.}
\label{fig:des}
\end{figure}

\mypara{Impact of Generated Prompt}
In hybrid detection with generated prompts, we rely on the BLIP model.
Here, we explore whether the quality of the BLIP-generated prompts affects the detection performance.
To measure the quality of the generated prompts by BLIP, we leverage a new term called prompt descriptiveness~\cite{SDTLH22, MPJ13, DMMPRRSYB22, EDC21}.
Prompt descriptiveness can be quantitatively measured by computing the cosine similarity between a prompt's embedding and its image's embedding generated by CLIP.\footnote{Note that the descriptiveness is the same as the one used in the previous analysis regarding why prompts can enhance detection performance.}
Such similarity demonstrates the degree of match between the generated prompts and the given images.
\autoref{fig:des} depicts the relation between the detection performance and the descriptiveness of the generated prompts.
We can see that in general, higher descriptiveness leads to better detection performance.
Also, after a certain descriptiveness value, the detection performance becomes stable across all models and datasets.
This shows the robustness of using BLIP-generated prompts in our hybrid detector.

\begin{figure*}[!t]
\centering
\begin{subfigure}{0.5\columnwidth}
\includegraphics[width=\columnwidth]{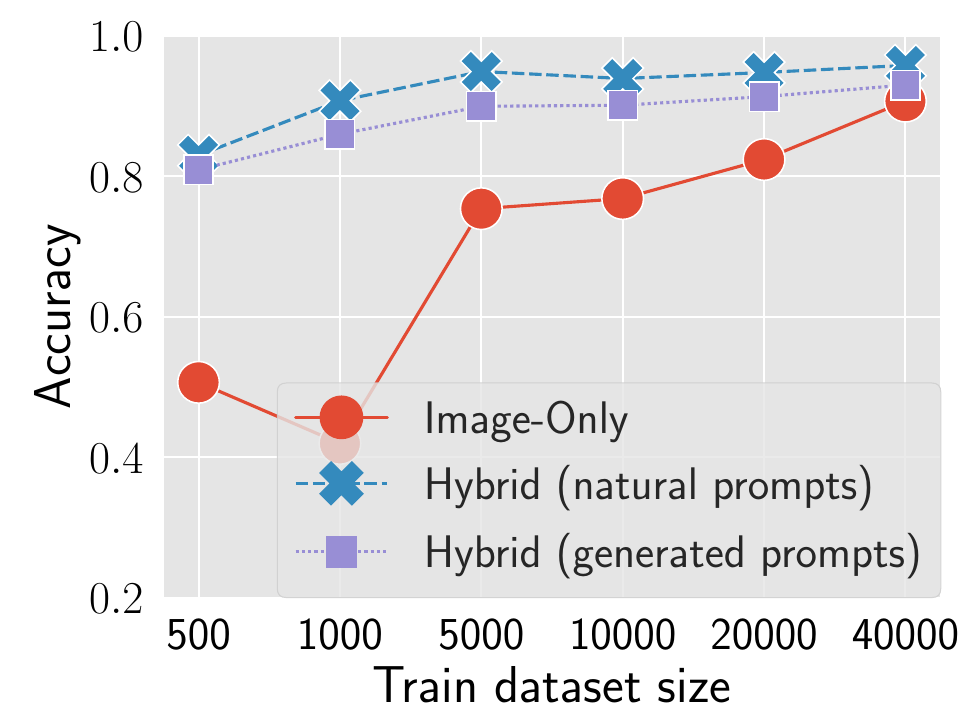}
\caption{SD}
\label{fig:size_stable}
\end{subfigure}
\begin{subfigure}{0.5\columnwidth}
\includegraphics[width=\columnwidth]{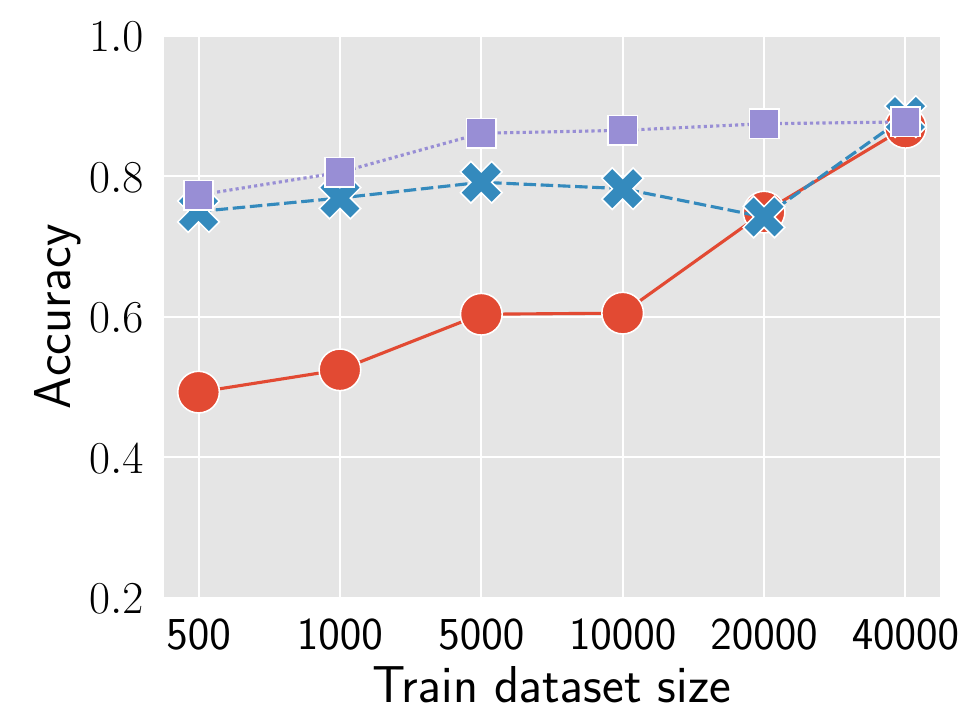}
\caption{LD}
\label{fig:size_ldm}
\end{subfigure}
\begin{subfigure}{0.5\columnwidth}
\includegraphics[width=\columnwidth]{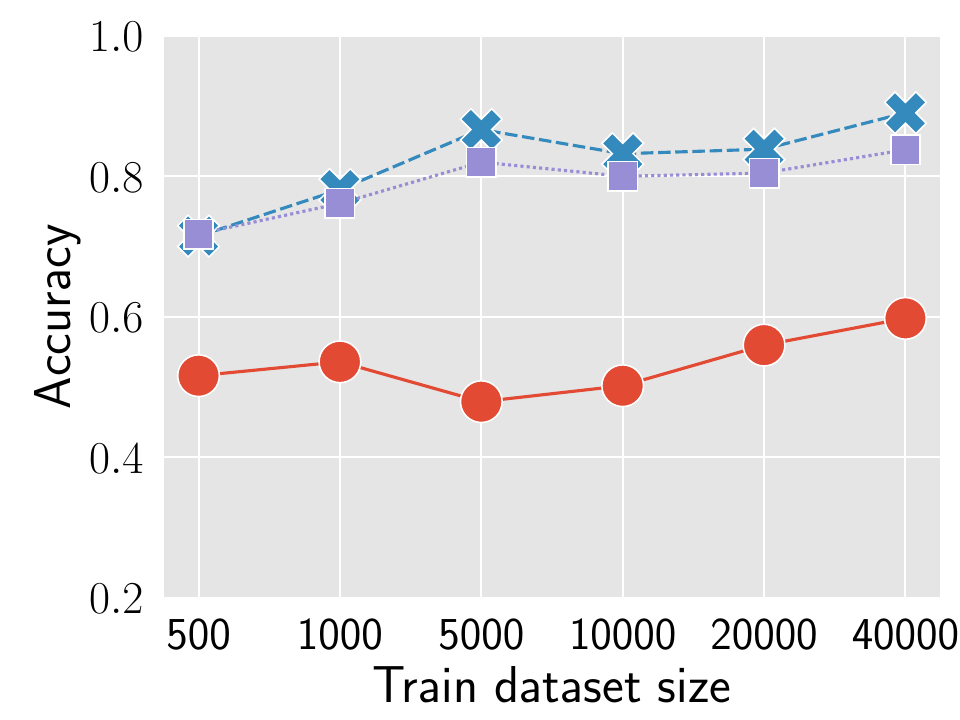}
\caption{GLIDE}
\label{fig:size_glide}
\end{subfigure}
\begin{subfigure}{0.5\columnwidth}
\includegraphics[width=\columnwidth]{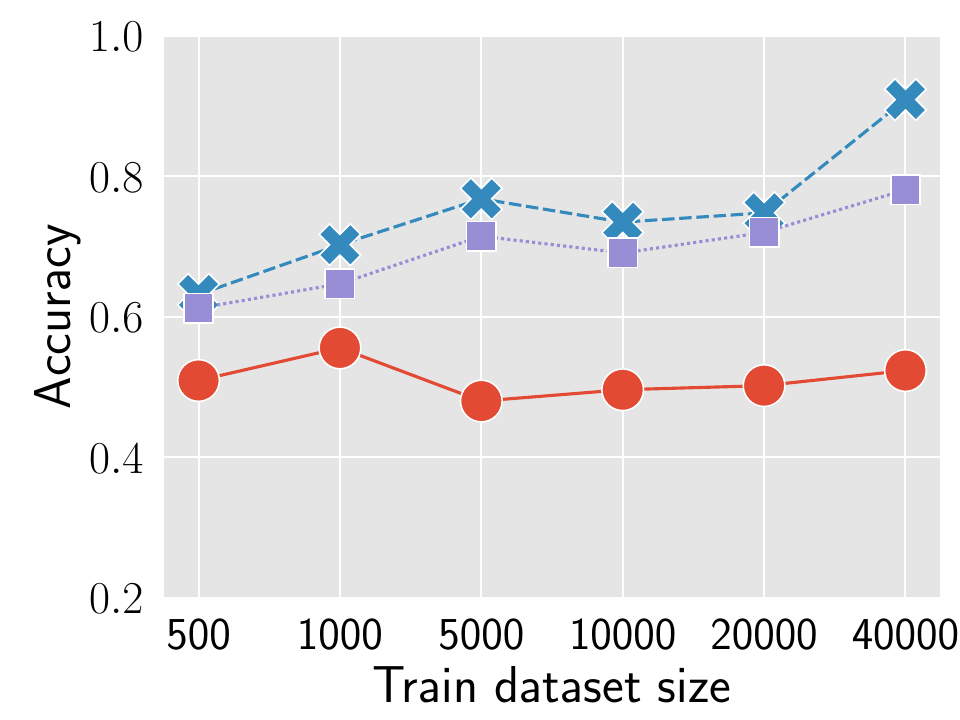}
\caption{\dalle}
\label{fig:size_dalle}
\end{subfigure}
\caption{The performance of detectors in terms of the training dataset size on SD+MSCOCO.
We conduct the evaluation on (a) SD+MSCOCO, (b) LD+MSCOCO, (c) GLIDE+MSCOCO, and (d) \dalle\!\!+MSCOCO.}
\label{fig:size}
\end{figure*}

\mypara{Impact of Training Dataset Size}
In this section, we explore the impact of the training dataset's size on the performance of our proposed fake image detection.
More concretely, for each text-to-image generation model, we train the detector on fake images from SD+MSCOCO by varying the size of the training dataset from 500 to 40,000 (half is real, half is fake).
Note that the default size we use in the previous evaluation is 40,000.

We report the detection performance in terms of the training dataset size in \autoref{fig:size}.
As expected, the performance of different detectors is indeed affected by the size of the training dataset, and the general trend is that all the detectors perform better with the increase in the training dataset size.
For instance, as shown in \autoref{fig:size_ldm}, when the training dataset size is 1,000, the hybrid detector can achieve an accuracy of 0.792 while the accuracy can be improved to 0.885 when the training dataset size is 40,000.
More encouragingly, we can also find that the hybrid detector achieves strong performance even with a small training dataset of only 500 images, which is much fewer than 40,000 images.
For example, in \autoref{fig:size_stable}, the hybrid detector achieves a high accuracy of 0.830 with only 500 training images.
Finally, we again find that the hybrid detector performs much better than the image-only detector, even with different sizes of the training dataset.
For example, in \autoref{fig:size_glide}, the hybrid detector achieves an accuracy of 0.714 with 500 training images, while the image-only detector achieves only 0.523 with 40,000 training images.
These results again demonstrate that introducing prompts together with images is beneficial to differentiate between fake and real images.

%-------------------------------------------------------------------------------
\subsection{Takeaways}
%-------------------------------------------------------------------------------

In summary, to answer \textbf{RQ1}, we propose fake image detection by training a binary detector to differentiate fake images generated by text-to-image generation models from real images.
Specially, we propose two methods to construct the binary detector, namely image-only and hybrid.
Our evaluation shows that the fake images from various models can indeed be differentiated from the real ones.
Moreover, the hybrid detector can obtain much better performance compared to the image-only detector, which demonstrates that introducing prompts together with images can indeed amplify the differences between fake and real images.

%-------------------------------------------------------------------------------
\section{Fake Image Attribution}
\label{section:source}
%-------------------------------------------------------------------------------

The previous section has shown that fake image detection, especially the hybrid detection we have proposed, can achieve remarkable performance.
In this section, we explore whether fake images generated by various text-to-image generation models can be attributed to their source models, i.e., fake image attribution.
We start by introducing our design goals.
We then describe how to construct the fake image attributor.
Finally, we present the evaluation results.

%-------------------------------------------------------------------------------
\subsection{Design Goals}
%-------------------------------------------------------------------------------

To attribute fake images to their source models, we follow two design goals.
\begin{itemize}
\item \textbf{Tracking Sources of Fake Images.}
The primary goal of fake image attribution is to effectively attribute different fake images to their source generation models.
The aim of attribution is to let a model owner be held responsible for the model's (possible) misuse.
Previously, fake image attribution has been studied in the context of traditional generation models, like GANs~\cite{YDF19}.
\item \textbf{Agnostic to Datasets.}
In the real world, a fake image can be generated by a text-to-image generation model  based on a prompt from any distribution.
Therefore, to be more practical, the attribution should be independent of the prompt distribution.
\end{itemize}

%-------------------------------------------------------------------------------
\subsection{Methodology}
%-------------------------------------------------------------------------------

To attribute the fake images to their sources, we construct fake image attribution by training a multi-class classifier, referred to as an attributor, with each class corresponding to one model.
As aforementioned, the attributor should be agnostic to datasets; thus, we establish the multi-class classifier based on prompts from only one dataset, e.g., MSCOCO, and test it on prompts from other datasets like Flickr30k.

Similar to fake image detection, we propose two different approaches to establish the attributor, namely image-only and hybrid.
The image-only attributor accepts only images as input, and the hybrid attributor accepts both images and their corresponding prompts as input.

\mypara{Image-Only Attribution}
The process of establishing our image-only attributor can also be divided into three stages, namely, data collection, dataset construction, and attributor construction.

\begin{itemize}
\item \mypara{Data Collection} 
We first randomly sample 20,000 images from MSCOCO as real images. 
Then, we use the prompts of these 20,000 images to query each model to get 20,000 fake images accordingly. 
Here, we adopt SD, LD, and GLIDE to generate fake images.
In total, we have obtained 60,000 fake images.
The reason we do not consider \dalle is that we will use \dalle for the experiments regarding adaptation to other models (see \autoref{section:source-discussion}).
\item \mypara{Dataset Construction} 
We label all real images as 0 and all fake images from the same model as the same class.
Concretely, we label the fake images from SD/LD/GLIDE as 1/2/3.
We then create a training dataset containing a total of 80,000 images with four classes.
\item \mypara{Attributor Construction}
We build the fake image attributor, i.e., a multi-class classifier, that accepts images as input and outputs the multi-class prediction, i.e., 0-real, 1-SD, 2-LD, or 3-GLIDE.
We train the attributor from scratch using the created training dataset in conjunction with classical training techniques.
Similar to the fake image detector, we leverage ResNet18~\cite{HZRS16} as the attributor's architecture.
\end{itemize}
After we have trained the attributor, we evaluate the performance of the trained attributor for attributing images from various sources (i.e., real, SD, LD, and GLIDE) given prompts from the other dataset (i.e., Flickr30k).
For testing the attributor, we sample the same number of images for all four classes, i.e., 10,000 each and 40,000 in total.

\mypara{Hybrid Attribution}
The previous evaluation in fake-image detection has demonstrated the superior performance of the hybrid detector, verifying that introducing prompts together with images can amplify the differences between fake and real images.
We now conduct the study to investigate whether a similar enhancement can be observed in the case of hybrid attribution.

The hybrid attributor is quite similar to the above image-only attributor, which also consists of three stages, i.e., data collection, dataset construction, and attributor construction.
\begin{itemize}
\item \mypara{Data Collection} 
To collect the images from various sources, we follow the same step as the first step for the image-only attributor.
\item \mypara{Dataset Construction} 
Since our hybrid attributor takes images and prompts as input, we label all real images with their corresponding prompts as 0 and all fake images from the same model with their corresponding prompts as the same class.
Similarly, we then create a training dataset containing a total of 80,000 prompt-images pairs with four classes.
\item \mypara{Attributor Construction} To exploit the prompt information, we again use CLIP's image encoder and text encoder as feature extractors to obtain high-level embeddings of images and prompts.
Then, we concatenate image embeddings and text embeddings together as new embeddings and use these embeddings to train a multi-class classifier, which is also a 2-layer multilayer perceptron, as our attributor.
\end{itemize}
In order to evaluate the trained hybrid attributor, we need images and their corresponding prompts.
We again consider two scenarios here, one in which we can directly obtain prompts for the images from the dataset and the other in which we can only generate prompts for the images relying on BLIP.

%-------------------------------------------------------------------------------
\subsection{Results}
%-------------------------------------------------------------------------------

In this section, we present the performance of our proposed two types of fake image attribution.

\mypara{Image-Only Attribution}
We report the performance of image-only attribution in \autoref{table:text-track}.
Note that the random guess for the 4-class classification task is only 0.25.
We can find that our proposed image-only attributor can achieve remarkable performance.
For instance, the image-only attributor can achieve an accuracy of 0.864 on images from various sources given the prompts sampled from MSCOCO.
These results indicate that the fake images can be effectively attributed to their corresponding text-to-image generation models.
We further show the unique feature extracted from each model in \autoref{section:source-discussion}, which implies that different models may leave unique fingerprints in the fake images they generate.

\begin{figure*}[!t]
\centering
\begin{subfigure}{0.5\columnwidth}
\includegraphics[width=\columnwidth]{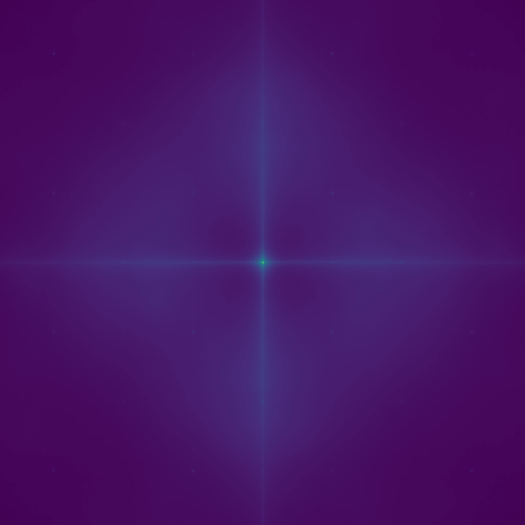}
\caption{SD}
\label{fig:fig_stable}
\end{subfigure}
\begin{subfigure}{0.5\columnwidth}
\includegraphics[width=\columnwidth]{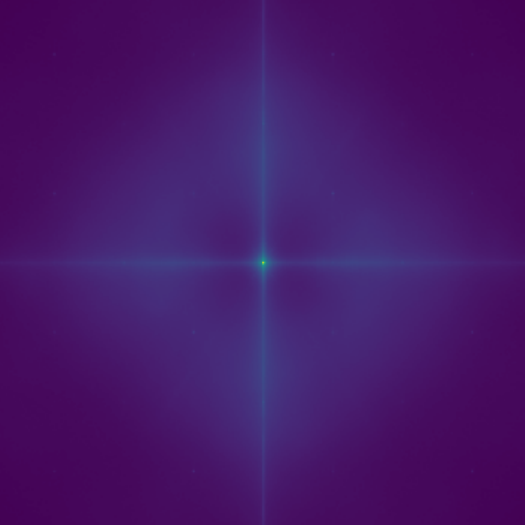}
\caption{LD}
\label{fig:fig_ldm}
\end{subfigure}
\begin{subfigure}{0.5\columnwidth}
\includegraphics[width=\columnwidth]{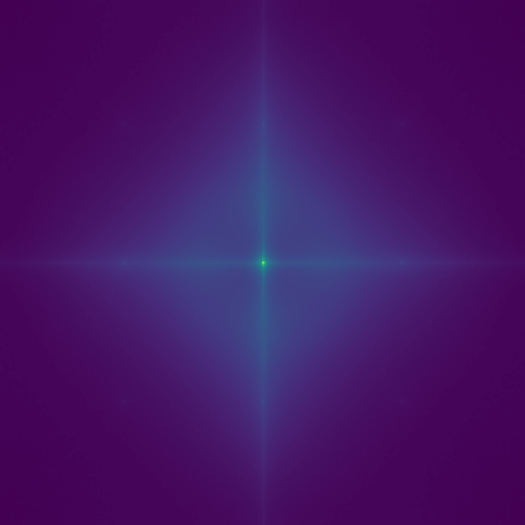}
\caption{GLIDE}
\label{fig:fig_glide}
\end{subfigure}
\begin{subfigure}{0.5\columnwidth}
\includegraphics[width=\columnwidth]{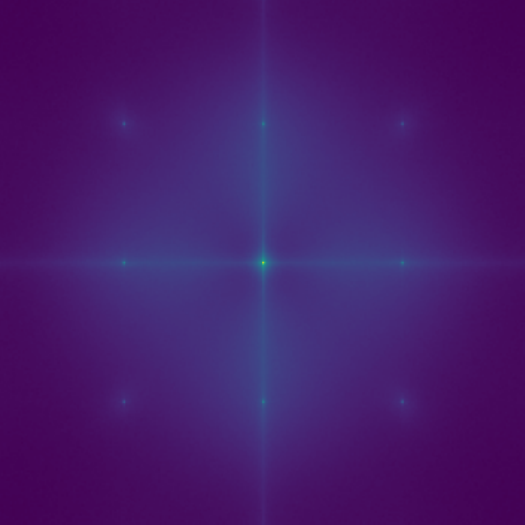}
\caption{\dalle}
\label{fig:fig_dalle}
\end{subfigure}
\caption{The visualization of frequency analysis on fake images generated by (a) SD, (b) LD, (c) GLIDE, and (d) \dalle.}
\label{fig:visual_freq}
\end{figure*}

Further, the image-only attributor can also achieve a high accuracy of 0.863 on images from various source models given the prompts sampled from the other dataset Flickr30k.
Note that we construct the attributor based on MSCOCO only.
This result indicates that our proposed image-only attribution is agnostic to datasets.

\begin{table}[!t]
\centering
\caption{The performance of image-only attributors and hybrid attributors.}
\label{table:text-track}
\begin{tabular}{lcc}
\toprule
~  & MSCOCO & Flickr30k \\
\midrule
Image-Only &  0.864 & 0.863 \\
Hybrid (natural prompts) &  0.936 & 0.933 \\
Hybrid (generated prompts) & 0.903 & 0.892 \\
\bottomrule
\end{tabular}
\end{table}

\mypara{Hybrid Attribution}
\autoref{table:text-track} also depicts the performance of our proposed hybrid attribution.
We can clearly see that hybrid attribution, no matter with or without natural prompts, achieves better performance than image-only attribution regardless of the dataset.
These results demonstrate once again that fake images can be successfully attributed to their corresponding text-to-image generation models.
Also, they verify that using prompts as an extra signal can improve attribution performance.

%-------------------------------------------------------------------------------
\subsection{Discussion}
\label{section:source-discussion}
%-------------------------------------------------------------------------------

The above evaluation demonstrates the effectiveness of our fake image attribution. 
We conjecture that each text-to-image generation model leaves a unique fingerprint in the fake images it generates.
Next, we verify this conjecture by visualizing the fingerprints of different models.
Besides, in the previous evaluation, the training and testing images for our attributor are disjoint but generated by the same set of text-to-image generation models.
We further explore how to adapt our attributor to other models that are not considered during training.

\mypara{Fingerprint Visualization}
Similar to visualizing the shared artifact across fake images (see \autoref{section:detection-disscussion}), we also draw the frequency spectra of different text-to-image generation models built on MSCOCO.
For each text-to-image generation model, we randomly select 2,000 fake images and then calculate the average of their Fourier transform outputs.

As shown in \autoref{fig:visual_freq}, we can clearly observe that there are distinct patterns in images generated by different text-to-image generation models, especially in GLIDE and \dalle.
We can also find that the frequency spectra of SD is similar to that of LD, which can explain why the image-only detector built on SD can also achieve very strong performance on LD (see \autoref{fig:text}).
The reason behind this is that SD and LD follow similar algorithms, although trained on different datasets and different model architectures.
In conclusion, the qualitative evaluation verifies that each text-to-image generation model has its unique fingerprint.

\mypara{Adaptation to Unseen Models}
In previous experiments, we evaluate attribution on fake images generated by models considered during training.
However, there are instances when we encounter fake images that are not from models involved in training, i.e., unseen models.
Next, we explore how to adapt our attributor to unseen models.

\begin{figure}[!t]
\centering
\begin{subfigure}{0.49\columnwidth}
\includegraphics[width=\columnwidth]{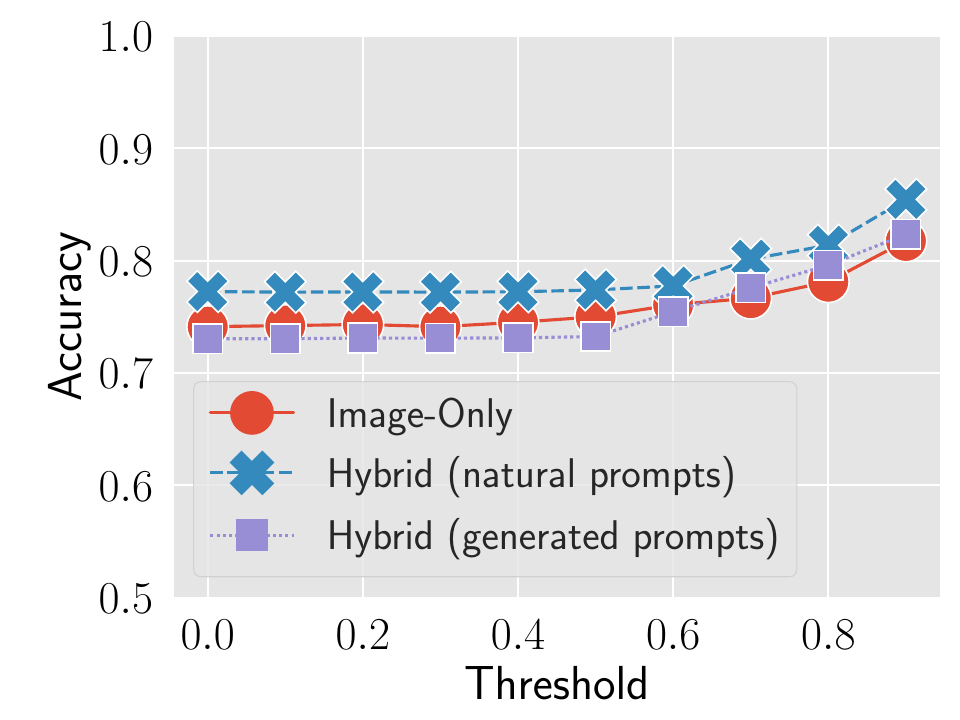}
\caption{MSCOCO}
\label{fig:threshold_coco}
\end{subfigure}
\begin{subfigure}{0.49\columnwidth}
\includegraphics[width=\columnwidth]{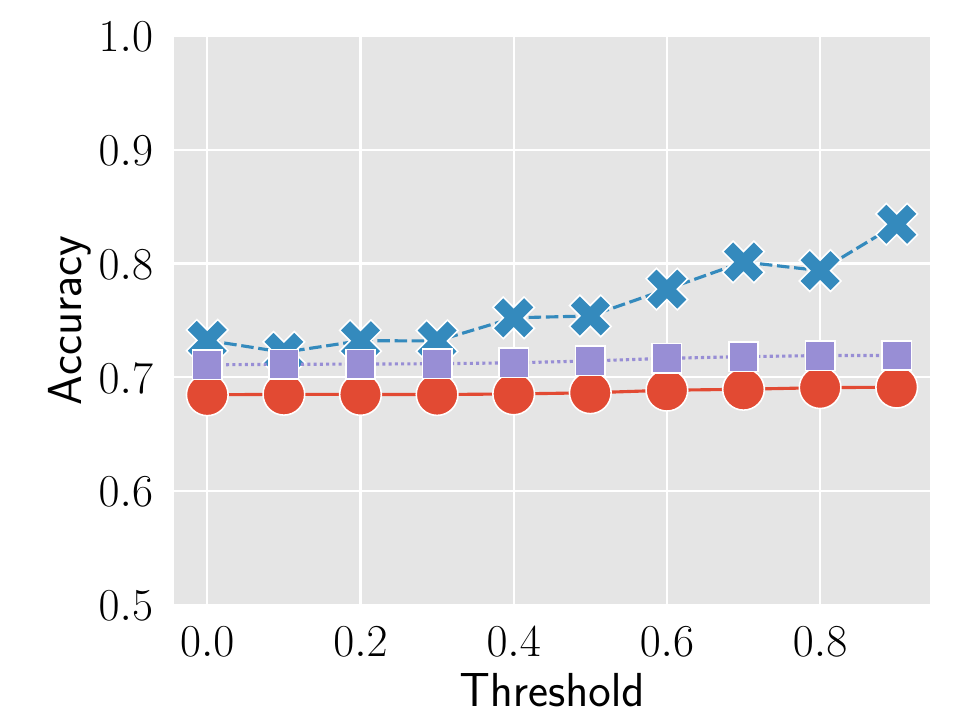}
\caption{Flickr30k}
\label{fig:threshold_flickr}
\end{subfigure}
\caption{The performance of attributors on an unseen dataset \dalle in terms of different thresholds.
We conduct the evaluation on (a) MSCOCO and (b) Flickr30k.}
\label{fig:threshold}
\end{figure}

To this end, we propose a simple yet effective approach named confidence-based attribution.
The key idea is to attribute the unconfident samples from the attributor prediction, i.e., lower than a pre-defined threshold, to unseen models.
Here, all unseen models are considered as one class.\footnote{In the current version, our approach cannot differentiate fake images from multiple unseen text-to-image generation models.
We will leave this as future work.}
In our evaluation, we treat \dalle as one unseen model (as mentioned before).
To find a suitable threshold, we have experimented with values from 0 to 1 in a step of 0.1.
Note that here we extend the evaluation from four classes to five: real, SD, LD, GLIDE, and unseen; the testing dataset is still balanced.
Also, the attributor remains unchanged, i.e., it is still a 4-class classifier.
\autoref{fig:threshold} shows that both image-only and hybrid attributors can achieve good performance in all cases.
Encouragingly, the 0.9 threshold can lead to the best attribution performance.
Moreover, we can still conclude that hybrid attribution can achieve better performance than image-only attribution in both settings.
These results indicate that with a simple modification, our attribution can be adapted to unseen models.

%-------------------------------------------------------------------------------
\subsection{Ablation Study}
%-------------------------------------------------------------------------------

\begin{figure}[!t]
\centering
\begin{subfigure}{0.49\columnwidth}
\includegraphics[width=\columnwidth]{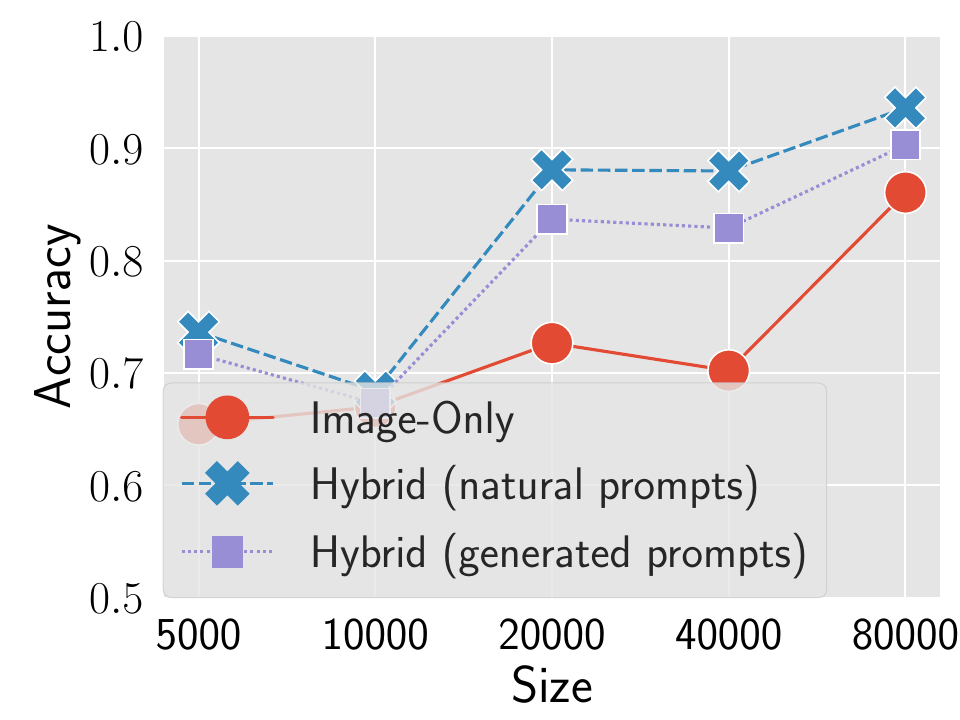}
\caption{MSCOCO}
\label{fig:model_level_size_mscoco}
\end{subfigure}
\begin{subfigure}{0.49\columnwidth}
\includegraphics[width=\columnwidth]{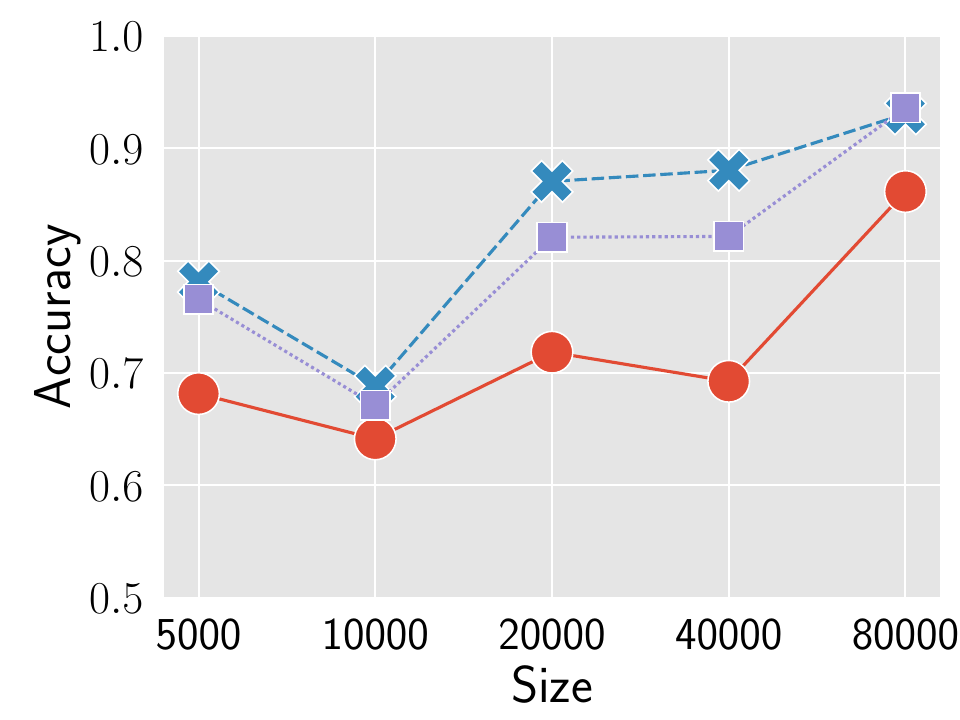}
\caption{Flickr30k}
\label{fig:model_level_size_flickr}
\end{subfigure}
\caption{The performance of attributors in terms of the training dataset size on MSCOCO.
We conduct the evaluation on (a) MSCOCO and (b) Flickr30k.}
\label{fig:model_level_size}
\end{figure}

\mypara{Impact of Training Dataset Size}
Here, we explore the effect of the training data size on attribution performance.
The experimental results are depicted in \autoref{fig:model_level_size}.
We can see that the size of training data indeed has a great influence on attribution performance. 
For example, when the training dataset size is 5,000, the hybrid attributor can achieve an accuracy of 0.736, while the accuracy can be improved to 0.946 when the training dataset size increases to 80,000.
Besides, we can find that the hybrid attributor requires less data to achieve a stable performance compared to the image-only attributor.
For example, hybrid attribution achieves a huge performance improvement from 10,000 to 20,000 in training dataset size, while for image-only attribution, a similar improvement happens when the training dataset size increases from 40,000 to 80,000.
From this phenomenon, we can conclude that hybrid attribution achieves good performance even with a small amount of training data.

%-------------------------------------------------------------------------------
\subsection{Takeaways}
%-------------------------------------------------------------------------------

In summary, to answer \textbf{RQ2}, we propose image-only attribution and hybrid attribution to track the source of fake images.
Empirical results indicate that fake images can be successfully attributed to their sources.
We further conduct a qualitative analysis that verifies the existence of unique fingerprints left by different text-to-image generation models in their generated images. 
Also, we show that our method can be easily adapted to other unseen models.

%-------------------------------------------------------------------------------
\section{Prompt Analysis}
%-------------------------------------------------------------------------------

One of the major differences between text-to-image generation models and traditional generation models is that the former takes a prompt as input.
In this section, we investigate which kinds of prompts are more likely to generate authentic images (\textbf{RQ3}).
To answer this question, we perform a comprehensive prompt analysis from semantic and structural perspectives.

\begin{table*}[!htbp]
\centering
\caption{Top five prompts which can generate the most real or fake images, determined by the image-only detector.
Gray cells in Real mean the prompt mainly describe the details of the subject.
Gray cells in Fake mean the prompt mainly describe the environment where the subject is located.}        
\label{table:case_study}
\tabcolsep 8pt
\begin{tabular}{c|cc|cc}
\toprule
Rank & \multicolumn{2}{c|}{Real}  & \multicolumn{2}{c}{Fake}  \\
\midrule
Top1 & 
\begin{minipage}[b]{0.2\columnwidth}
\centering
\raisebox{-.5\height}{\includegraphics[width=\linewidth]{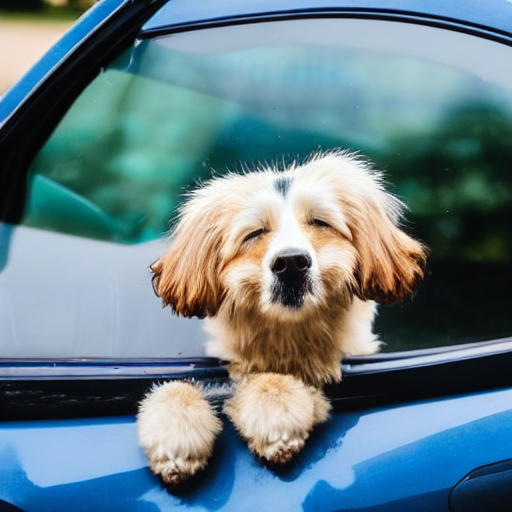}}
\end{minipage}
&
\cellcolor[HTML]{E6E6E6}
\begin{tabular}[c]{@{}l@{}}A dog hanging out of \\ a side window on a car\end{tabular}
&
\begin{minipage}[b]{0.2\columnwidth}
\centering
\raisebox{-.5\height}{\includegraphics[width=\linewidth]{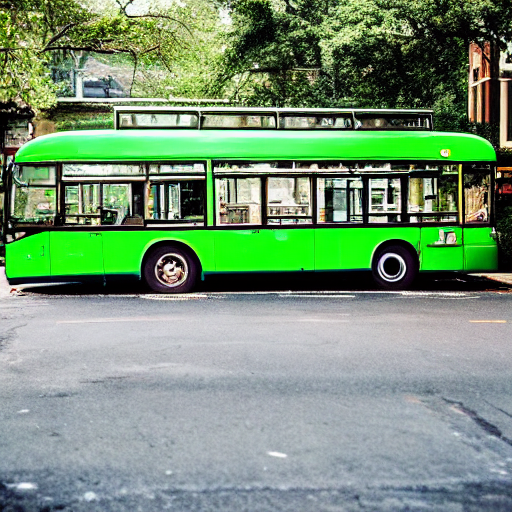}}
\end{minipage}
 &
 \cellcolor[HTML]{E6E6E6}
 \begin{tabular}[c]{@{}l@{}}A green bus is parked \\ on the side of the street\end{tabular}
 \\
\hline
Top2 & 
\begin{minipage}[b]{0.2\columnwidth}
\centering
\raisebox{-.5\height}{\includegraphics[width=\linewidth]{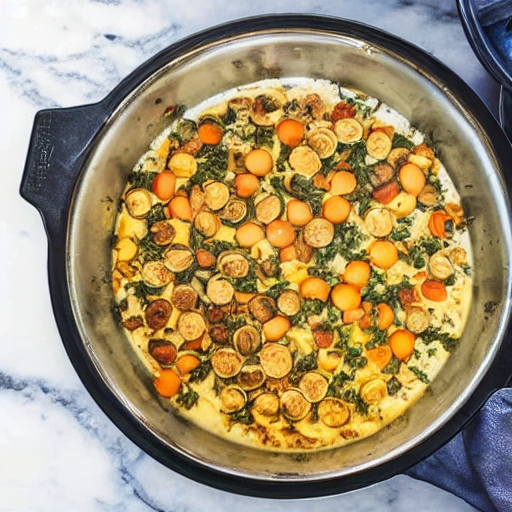}}
\end{minipage}
&
\cellcolor[HTML]{E6E6E6}
\begin{tabular}[c]{@{}l@{}}A pan filled with food \\ sitting on a stove top \end{tabular}
 &
\begin{minipage}[b]{0.2\columnwidth}
\centering
\raisebox{-.5\height}{\includegraphics[width=\linewidth]{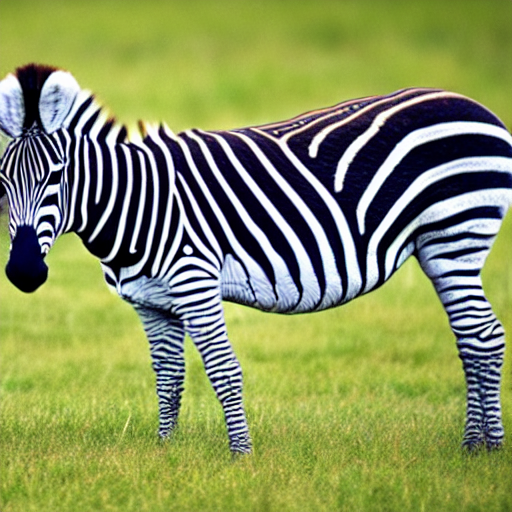}}
\end{minipage}
 &
\cellcolor[HTML]{E6E6E6} 
\begin{tabular}[c]{@{}l@{}}THERE IS A ZEBRA THAT IS \\ EATING GRASS IN THE YARD  \end{tabular}
\\
\hline
Top3 & 
\begin{minipage}[b]{0.2\columnwidth}
\centering
\raisebox{-.5\height}{\includegraphics[width=\linewidth]{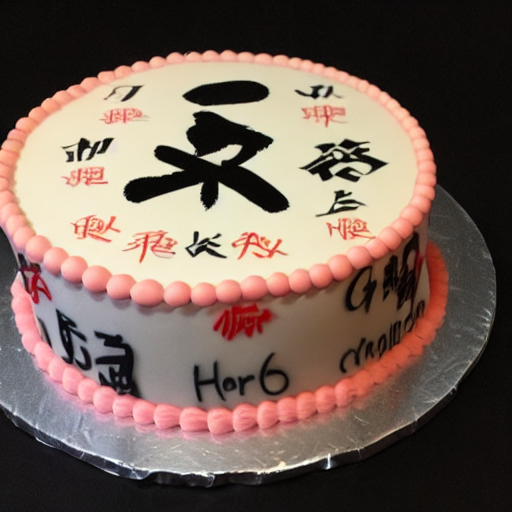}}
\end{minipage}
 &
\cellcolor[HTML]{E6E6E6}
\begin{tabular}[c]{@{}l@{}}A birthday cake with English \\ and Chinese characters \end{tabular}
  &
\begin{minipage}[b]{0.2\columnwidth}
\centering
\raisebox{-.5\height}{\includegraphics[width=\linewidth]{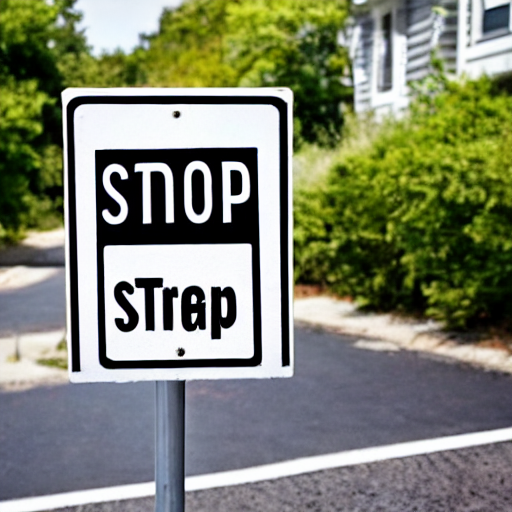}}
\end{minipage}
 &
 \begin{tabular}[c]{@{}l@{}}I sign that indicates the street \\ name posted above a stop sign \end{tabular}
 \\
\hline
Top4 & 
\begin{minipage}[b]{0.2\columnwidth}
\centering
\raisebox{-.5\height}{\includegraphics[width=\linewidth]{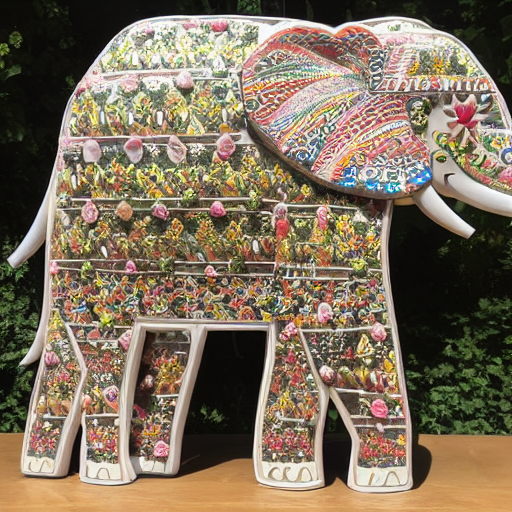}}
\end{minipage}
 &
  \begin{tabular}[c]{@{}l@{}}There is an elephant-shaped figure \\ next to other decorations  \end{tabular}

 & 
 \begin{minipage}[b]{0.2\columnwidth}
\centering
\raisebox{-.5\height}{\includegraphics[width=\linewidth]{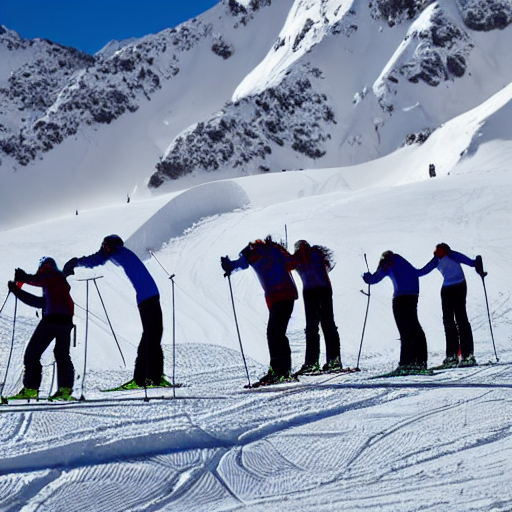}}
\end{minipage}
 &
\cellcolor[HTML]{E6E6E6} 
\begin{tabular}[c]{@{}l@{}}A group of skiers as they \\ ski on the snow \end{tabular}
\\
\hline
Top5 &
\begin{minipage}[b]{0.2\columnwidth}
\centering
\raisebox{-.5\height}{\includegraphics[width=\linewidth]{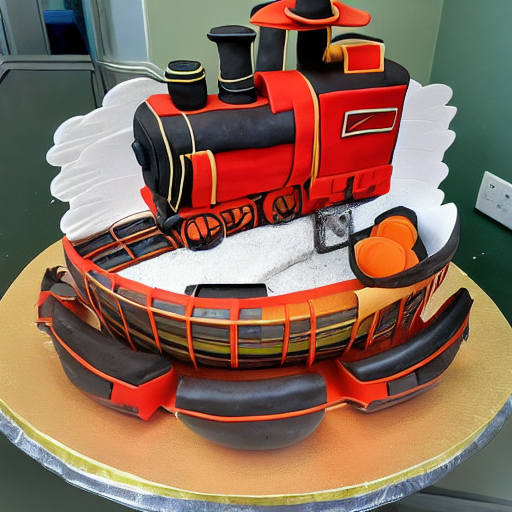}}
\end{minipage}
 &
\cellcolor[HTML]{E6E6E6} 
\begin{tabular}[c]{@{}l@{}}there is a cake and donuts \\ that look like a train \end{tabular}
 
 &
 \begin{minipage}[b]{0.2\columnwidth}
\centering
\raisebox{-.5\height}{\includegraphics[width=\linewidth]{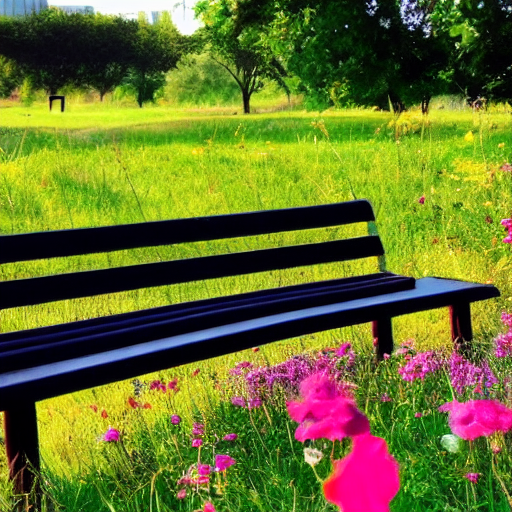}}
\end{minipage}
 &
 \cellcolor[HTML]{E6E6E6} 
 \begin{tabular}[c]{@{}l@{}} A bench is surrounded by \\ grass and a few flowers \end{tabular}
 \\
\bottomrule
\end{tabular}
\end{table*}

\begin{figure}[!t]
\centering
\includegraphics[width=\columnwidth]{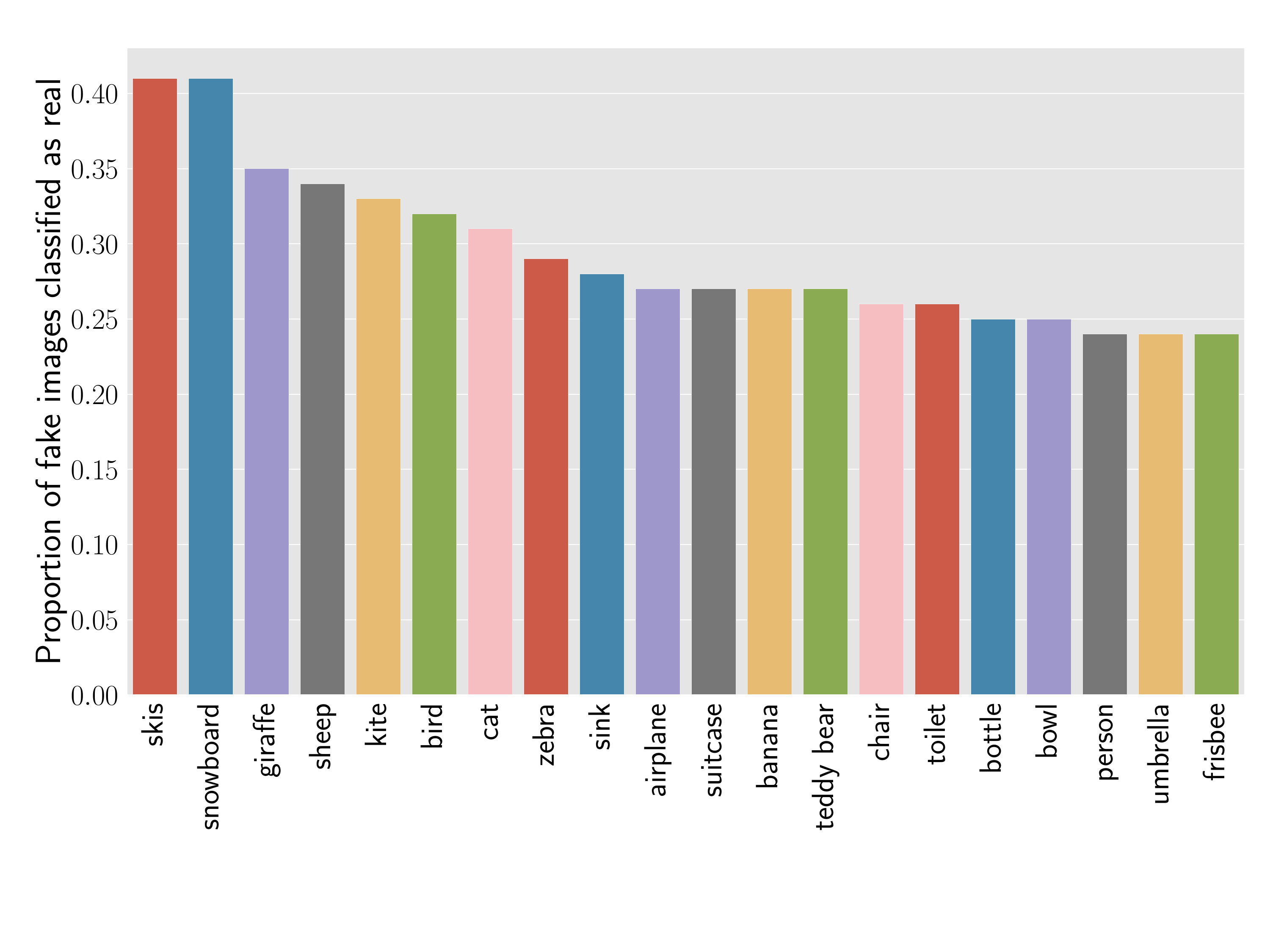}
\caption{The top twenty topics of prompts in terms of the proportion of the corresponding generated fake images being classified as real by the image-only detector.
The topics are extracted from the MSCOCO dataset.}
\label{fig:cate}
\end{figure}

\begin{figure}[!t]
\centering
\begin{subfigure}{0.47\columnwidth}
\includegraphics[width=\columnwidth]{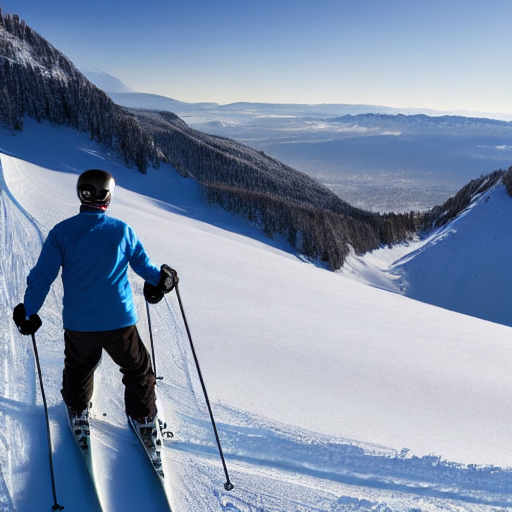}
\caption{Skis}
\label{fig:snowboard1}
\end{subfigure}
\begin{subfigure}{0.47\columnwidth}
\includegraphics[width=\columnwidth]{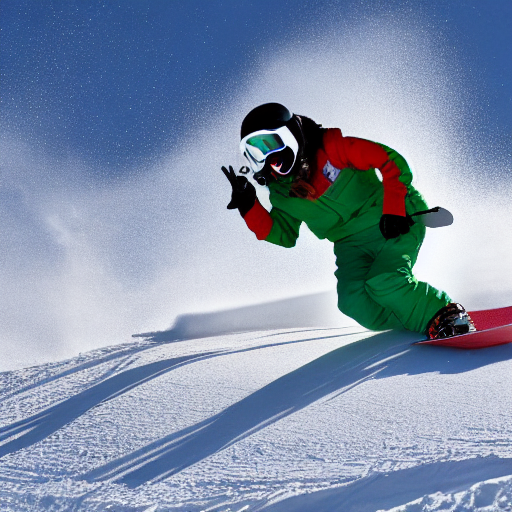}
\caption{Snowboard}
\label{fig:snowboard2}
\end{subfigure}
\caption{Examples of fake images generated by SD given prompts with topics ``skis'' and ``snowboard.''}
\label{fig:snowboard}
\end{figure}

%-------------------------------------------------------------------------------
\subsection{Semantics Analysis}
\label{section:topic}
%-------------------------------------------------------------------------------

We first conduct semantic analysis on prompts based on their topics.
Concretely, we group prompts into different clusters by topic.
Then, for each cluster/topic, we calculate the proportion of the corresponding fake images being classified as real images by our image-only detector (\autoref{section:universal_detection}).
A cluster with a higher proportion indicates the prompts with the underlying topic have a higher chance of generating authentic images. 
As we focus on the authenticity of an image itself, we adopt the image-only detector instead of the hybrid detector.
Note that our analysis is conducted on fake images generated by SD given prompts from MSCOCO.

We first utilize a straightforward method to group prompts relying on the topics provided by MSCOCO.
In total, there are 80 topics in MSCOCO.
We select the top twenty topics with the highest real image proportion decided by the image-only detector and report the results in \autoref{fig:cate}.
We can clearly observe that among the top twenty topics, ``skis'' and ``snowboard'' are ranked the highest.
Also, there are many topics related to animals, such ``sheep,'' ``cat,'' ``zebra,'' etc.

Though the topics from MSCOCO are straightforward, they may not be able to represent the full semantics of the images.
Therefore, we take another approach.
Specifically, we take advantage of sentence transformer~\cite{RG19} based on BERT~\cite{DCLT19} to generate embeddings for the prompts and then group the embeddings with DBSCAN~\cite{EKSX96}, an advanced clustering method.
The advantage of the second approach is that it can implicitly reflect the in-depth semantics of the prompts, which is also a common practice in the natural language processing literature.
However, the disadvantage of this approach is that the concrete topic of each cluster needs to be manually summarized.
By manually checking, the cluster with the highest real image proportion is related to the topic ``person.''
Ostensibly, this is different from the results of the first approach (``skis'' and ``snowboard'' ranked the highest), which is based on the topics provided by MSCOCO. 
However, by manually checking fake images by prompts with topics ``skis'' and ``snowboard,'' we discover that most of them depict ``person'' as well.
We show some examples in \autoref{fig:snowboard}.
This indicates the prompts related to ``person'' are likely to generate authentic fake images. 

We further extract the top 5 prompts that can generate the most real and fake images, respectively, according to the image-only detector, and list them in \autoref{table:case_study}.
We can also find that detailed descriptions of the subjects contribute to the generation of authentic images.
For example, of the top five real prompts, four provide a detailed description of the subject, while four of the top five fake prompts describe the environment where the subject is located, rather than the subject itself.
In the future, we plan to investigate in-depth the relationship between the prompts' semantics and the generated images' authenticity.

%-------------------------------------------------------------------------------
\subsection{Structure Analysis}
%-------------------------------------------------------------------------------

After semantic analysis, we now conduct the structure analysis.
Specifically, we study prompt structure from two angles, i.e., the length and the proportion of nouns.
The length of the prompt reflects the prompt complexity.
The proportion of nouns is related to the number of objects appearing in the fake image.
Here, we use Natural Language Toolkit (NLTK)~\cite{BL04} to compute the proportion of nouns in a prompt.

\begin{figure}[!t]
\centering
\begin{subfigure}{0.49\columnwidth}
\includegraphics[width=\columnwidth]{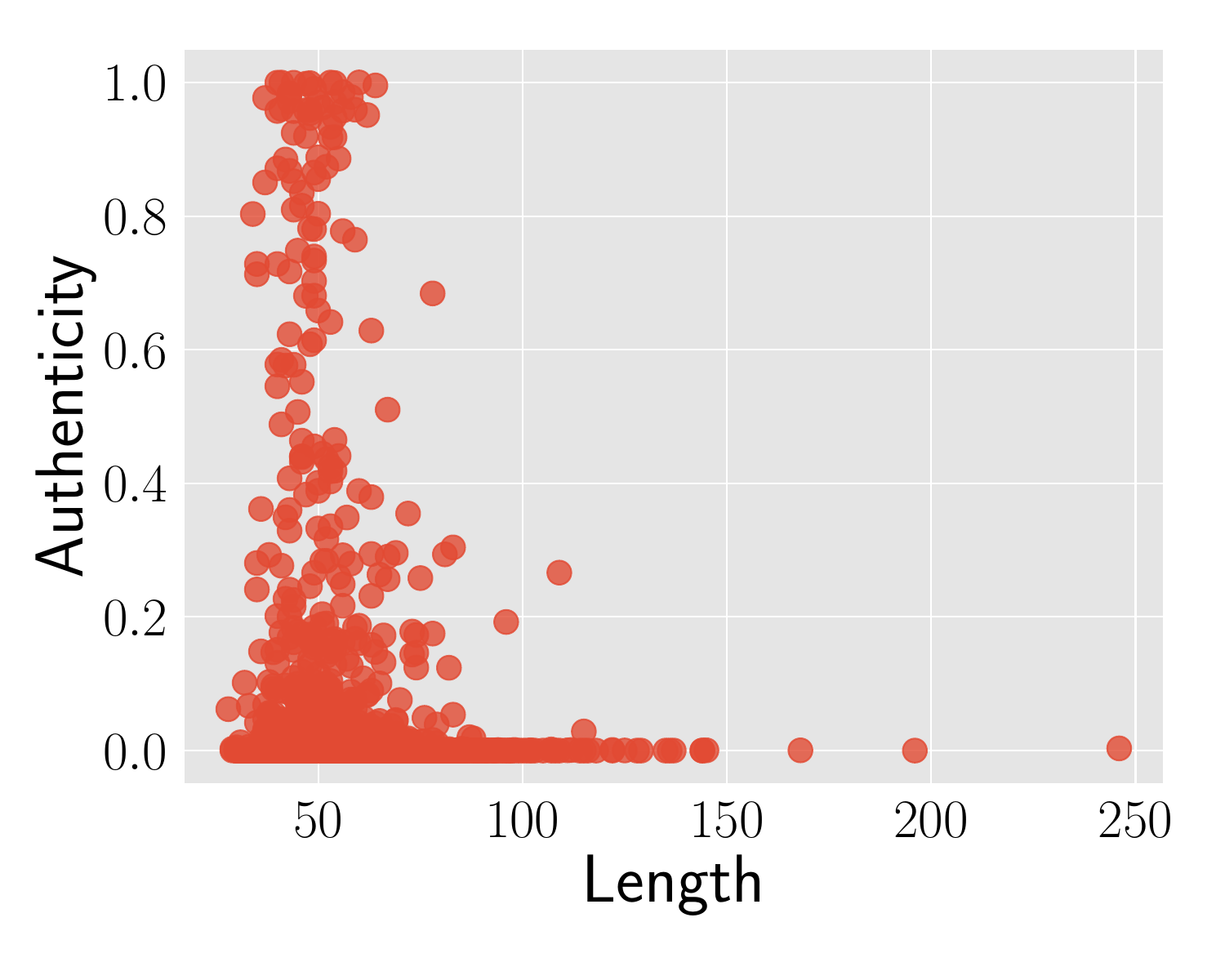}
\caption{Length}
\label{fig:prompt_len}
\end{subfigure}
\begin{subfigure}{0.49\columnwidth}
\includegraphics[width=\columnwidth]{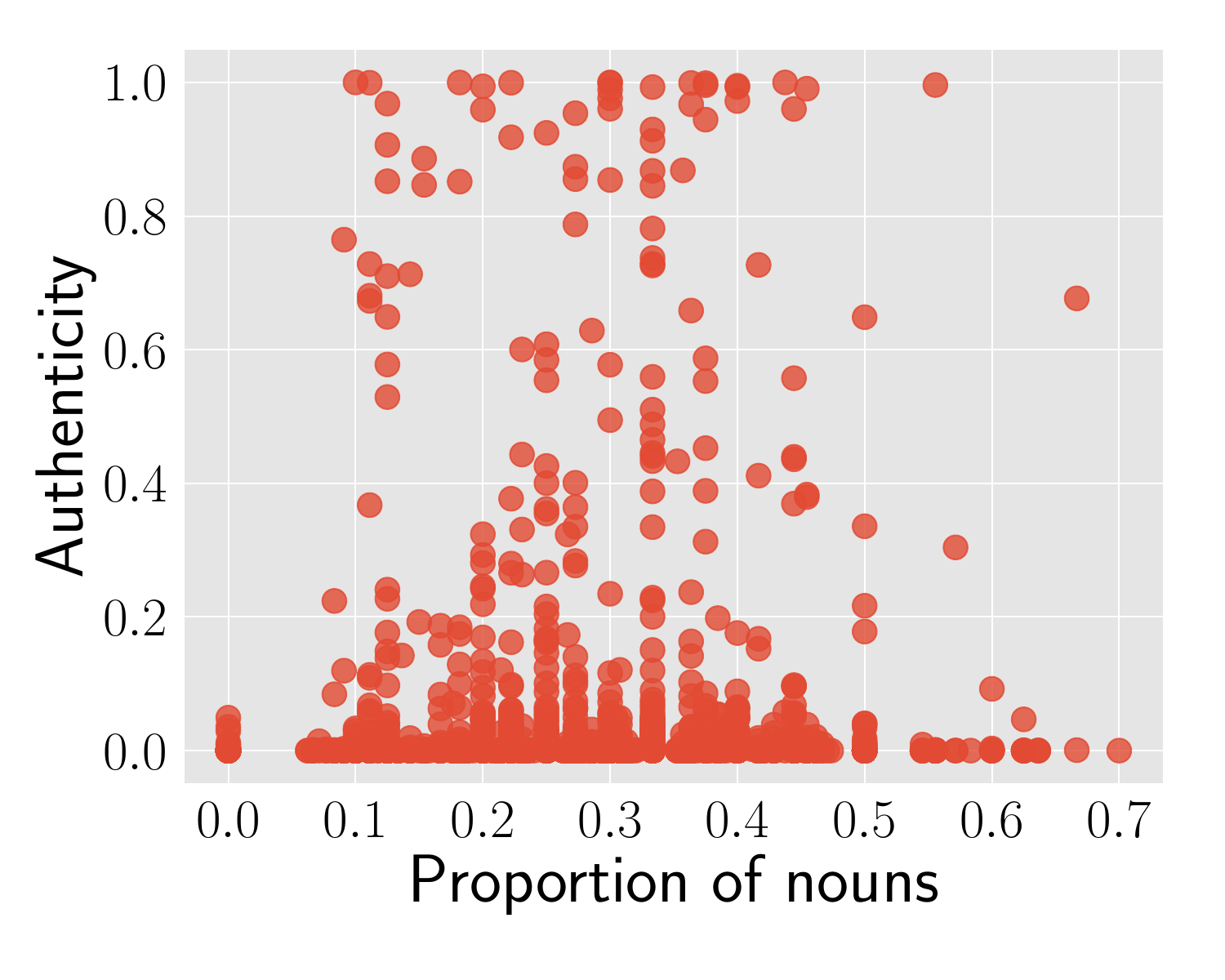}
\caption{Proportion of noun}
\label{fig:prompt_nn}
\end{subfigure}
\caption{The relationship between the length\textbackslash proportion of nouns in a prompt and the corresponding image's authenticity.}
\label{fig:structure}
\end{figure}

In our experiments, we randomly select 5,000 prompts from MSCOCO and then feed these prompts to SD to generate fake images.
Results are shown in \autoref{fig:structure}.
We can see from \autoref{fig:prompt_len} that both extremely long and short prompts cannot generate authentic images.
In addition, almost all high-authenticity images are generated by prompts with lengths between 25 to 75.
On the other hand, \autoref{fig:prompt_nn} shows that the proportion of nouns in prompts does not have a significant impact on fake images' authenticity.

%-------------------------------------------------------------------------------
\subsection{Takeaways}
%-------------------------------------------------------------------------------

In summary, we conduct semantic analysis and structure analysis to study which types of prompts are more likely to drive text-to-image generation models to generate fake images with high authenticity.
Empirical results demonstrate that a prompt with the topic ``person'' or length between 25 and 75 is more likely to produce authentic images, thus leading to difficulties in detection by our designed detectors.

%-------------------------------------------------------------------------------
\section{Related Work}
%-------------------------------------------------------------------------------

%-------------------------------------------------------------------------------
\subsection{Text-to-Image Generation}
%-------------------------------------------------------------------------------

Typically, text-to-image generation takes a text description (i.e., a prompt) as input and outputs an image that matches the text description.
Some pioneer works of text-to-image generation~\cite{RAYLSL16, ZXL17} are based on GANs~\cite{GPMXWOCB14}.
By combining a prompt embedding and a latent vector, the authors expect the GANs to generate an image depicting the prompt.
These works have stimulated more researchers~\cite{BHC18, LHPDDF19, WQWHC20, ZKBLY21, SWR20, LZZHHLG19} to study text-to-image generation models based on GANs, but using GANs does not always achieve good generation performance~\cite{RPGGVRCS21, RBLEO22}.

Recently, text-to-image generation has made great progress with the emergence of diffusion models~\cite{AT16, RBLEO22, NDRSMMSC21, SCSLWDGAMLSHFN22}.
Models in this domain normally take random noise and prompts as input and reduce noisy images to clear ones based on the guidance of prompts.
Currently, text-to-image generation based on diffusion models, such as DALL$\cdot$E~\cite{RPGGVRCS21}, Stable Diffusion~\cite{RBLEO22}, Imagen~\cite{SCSLWDGAMLSHFN22}, GLIDE~\cite{NDRSMMSC21} and \dalle~\cite{RDNCC22}, has achieved state-of-the-art performance compared to previous works. 
This is also the reason why we focus on such models in this work.

%-------------------------------------------------------------------------------
\subsection{Fake Image Detection and Attribution}
%-------------------------------------------------------------------------------

Wang et al.~\cite{WWZOE20} find that a simple CNN model can easily detect fake images generated by various types of traditional generation models (e.g., GANs~\cite{GPMXWOCB14} and low-level vision models~\cite{CCXK18, DCZXZ19}) from real images.
The authors argue that these fake images have some common defects that allow us to distinguish them from real images.
Yu et al.~\cite{YDF19} demonstrate that fake images generated by various traditional generation models can be attributed to their sources and reveal the fact that these traditional generation models leave fingerprints in the generated images.
Girish et al.~\cite{GSRS21} further propose a new attribution method to deal with the open-world scenario where the detector has no knowledge of the generation model.

We emphasize here that almost all existing works focus only on traditional generation models, such as GANs~\cite{GPMXWOCB14}, low-level vision models~\cite{CCXK18, DCZXZ19}, and perceptual loss generation models~\cite{CK17, LZM19}.
Detecting and attributing fake images generated by text-to-image generation models are largely unexplored.
In this work, we take the first step to systematically study the problem.

%-------------------------------------------------------------------------------
\section{Conclusion}
%-------------------------------------------------------------------------------

In this paper, we delve into three research questions concerning the detection and attribution of fake images generated by text-to-image generation models.
To solve the first research question of whether we can distinguish fake images apart from real ones, we propose fake image detection.
Our fake image detection consists of two types of detectors: an image-only detector and a hybrid detector.
The image-only detector utilizes images as the input to identify fake images, while the hybrid detector leverages both image information and the corresponding prompt information.
In the testing phase, if the hybrid detector cannot obtain the natural prompt of an image, we take advantage of BLIP, an image captioning model, to generate a prompt for the image.
Our extensive experiments show that while an image-only detector can achieve strong performance on certain text-to-image generation models, a hybrid detector can always have better performance.
These results demonstrate that fake images generated by different text-to-image generation models share common features. 
Also, prompts can serve as an extra ``anchor''  to help the detector better differentiate between fake and real images.

To tackle the second research question, we conduct the fake image attribution to attribute fake images from different text-to-image generation models to their source models.
Similarly, we develop two types of multi-class classifiers: an image-only attributor and a hybrid attributor. 
Empirical results show that both image-only attributor and hybrid attributor have good performance in all cases.
This implies that fake images generated by different text-to-image generation models enjoy different properties, which can also be viewed as fingerprints.

Finally, we address the third research question, i.e., which kinds of prompts are more likely to generate authentic images?
We study the properties of prompts from semantic and structural perspectives.
From the semantic perspective, we show that prompts with the topic ``person'' can achieve more authentic fake images compared to prompts with other topics.
From the structural perspective, our experiments reveal that prompts with lengths ranging from 25 to 75 allow text-to-image generation models to create more authentic fake images.

Overall, this work presents the first comprehensive study of detecting and attributing fake images  generated by state-of-the-art text-to-image generation models.
As our empirical results are encouraging, we believe our detectors and attributors can play an essential role in mitigating the threats caused by fake images created by the advanced generation models.
We will share our code to facilitate research in this field in the future.

%-----------------------------------------------------
\bibliographystyle{plain}
\begin{small}
\bibliography{normal_generated_py3} 

\begin{thebibliography}{10}

\bibitem{AT16}
James Atwood and Don Towsley.
\newblock {Diffusion-Convolutional Neural Networks}.
\newblock In {\em {Annual Conference on Neural Information Processing Systems
  (NIPS)}}, pages 1993--2001. NIPS, 2016.

\bibitem{BL04}
Steven Bird and Edward Loper.
\newblock {{NLTK:} The Natural Language Toolkit}.
\newblock In {\em {Annual Meeting of the Association for Computational
  Linguistics (ACL)}}. ACL, 2004.

\bibitem{BHC18}
Navaneeth Bodla, Gang Hua, and Rama Chellappa.
\newblock {Semi-supervised FusedGAN for Conditional Image Generation}.
\newblock In {\em {European Conference on Computer Vision (ECCV)}}, pages
  689--704. Springer, 2018.

\bibitem{CCXK18}
Chen Chen, Qifeng Chen, Jia Xu, and Vladlen Koltun.
\newblock {Learning to See in the Dark}.
\newblock In {\em {IEEE Conference on Computer Vision and Pattern Recognition
  (CVPR)}}, pages 3291--3330. IEEE, 2018.

\bibitem{CK17}
Qifeng Chen and Vladlen Koltun.
\newblock {Photographic Image Synthesis with Cascaded Refinement Networks}.
\newblock In {\em {IEEE International Conference on Computer Vision (ICCV)}},
  pages 1520--1529. IEEE, 2017.

\bibitem{CCKHKC18}
Yunjey Choi, Min{-}Je Choi, Munyoung Kim, Jung{-}Woo Ha, Sunghun Kim, and
  Jaegul Choo.
\newblock {StarGAN: Unified Generative Adversarial Networks for Multi-Domain
  Image-to-Image Translation}.
\newblock In {\em {IEEE Conference on Computer Vision and Pattern Recognition
  (CVPR)}}, pages 8789--8797. IEEE, 2018.

\bibitem{DCZXZ19}
Tao Dai, Jianrui Cai, Yongbing Zhang, Shu{-}Tao Xia, and Lei Zhang.
\newblock {Second-Order Attention Network for Single Image Super-Resolution}.
\newblock In {\em {IEEE Conference on Computer Vision and Pattern Recognition
  (CVPR)}}, pages 11065--11074. IEEE, 2019.

\bibitem{DCLT19}
Jacob Devlin, Ming{-}Wei Chang, Kenton Lee, and Kristina Toutanova.
\newblock {BERT: Pre-training of Deep Bidirectional Transformers for Language
  Understanding}.
\newblock In {\em {Conference of the North American Chapter of the Association
  for Computational Linguistics: Human Language Technologies (NAACL-HLT)}},
  pages 4171--4186. ACL, 2019.

\bibitem{DMMPRRSYB22}
Pierre~L. Dognin, Igor Melnyk, Youssef Mroueh, Inkit Padhi, Mattia Rigotti,
  Jarret Ross, Yair Schiff, Richard~A. Young, and Brian Belgodere.
\newblock {Image Captioning as an Assistive Technology: Lessons Learned from
  VizWiz 2020 Challenge}.
\newblock {\em {Journal of Artificial Intelligence Research}}, 2022.

\bibitem{EDC21}
Kreiss Elisa, Goodman~Noah D, and Potts Christopher.
\newblock {Concadia: Tackling Image Accessibility with Descriptive Texts and
  Context}.
\newblock {\em {CoRR abs/2104.08376}}, 2021.

\bibitem{EKSX96}
Martin Ester, Hans{-}Peter Kriegel, J{\"{o}}rg Sander, and Xiaowei Xu.
\newblock {A Density-Based Algorithm for Discovering Clusters in Large Spatial
  Databases with Noise}.
\newblock In {\em {International Conference on Knowledge Discovery and Data
  Mining (KDD)}}, pages 226--231. AAAI, 1996.

\bibitem{GSRS21}
Sharath Girish, Saksham Suri, Sai~Saketh Rambhatla, and Abhinav Shrivastava.
\newblock {Towards Discovery and Attribution of Open-World GAN Generated
  Images}.
\newblock In {\em {IEEE International Conference on Computer Vision (ICCV)}},
  pages 14094--14103. IEEE, 2021.

\bibitem{GPMXWOCB14}
Ian Goodfellow, Jean Pouget-Abadie, Mehdi Mirza, Bing Xu, David Warde-Farley,
  Sherjil Ozair, Aaron Courville, and Yoshua Bengio.
\newblock {Generative Adversarial Nets}.
\newblock In {\em {Annual Conference on Neural Information Processing Systems
  (NIPS)}}, pages 2672--2680. NIPS, 2014.

\bibitem{HZRS16}
Kaiming He, Xiangyu Zhang, Shaoqing Ren, and Jian Sun.
\newblock {Deep Residual Learning for Image Recognition}.
\newblock In {\em {IEEE Conference on Computer Vision and Pattern Recognition
  (CVPR)}}, pages 770--778. IEEE, 2016.

\bibitem{LHPDDF19}
Qicheng Lao, Mohammad Havaei, Ahmad Pesaranghader, Francis Dutil, Lisa
  Di{-}Jorio, and Thomas Fevens.
\newblock {Dual Adversarial Inference for Text-to-Image Synthesis}.
\newblock In {\em {IEEE International Conference on Computer Vision (ICCV)}},
  pages 7566--7575. IEEE, 2019.

\bibitem{LLXH22}
Junnan Li, Dongxu Li, Caiming Xiong, and Steven C.~H. Hoi.
\newblock {{BLIP:} Bootstrapping Language-Image Pre-training for Unified
  Vision-Language Understanding and Generation}.
\newblock {\em {CoRR abs/2201.12086}}, 2022.

\bibitem{LZM19}
Ke~Li, Tianhao Zhang, and Jitendra Malik.
\newblock {Diverse Image Synthesis From Semantic Layouts via Conditional
  {IMLE}}.
\newblock In {\em {IEEE International Conference on Computer Vision (ICCV)}},
  pages 4219--4228. IEEE, 2019.

\bibitem{LZZHHLG19}
Wenbo Li, Pengchuan Zhang, Lei Zhang, Qiuyuan Huang, Xiaodong He, Siwei Lyu,
  and Jianfeng Gao.
\newblock {Object-Driven Text-To-Image Synthesis via Adversarial Training}.
\newblock In {\em {IEEE Conference on Computer Vision and Pattern Recognition
  (CVPR)}}, pages 1274--12182. IEEE, 2019.

\bibitem{LMBHPRDZ14}
Tsung{-}Yi Lin, Michael Maire, Serge~J. Belongie, James Hays, Pietro Perona,
  Deva Ramanan, Piotr Doll{\'{a}}r, and C.~Lawrence Zitnick.
\newblock {Microsoft {COCO:} Common Objects in Context}.
\newblock In {\em {European Conference on Computer Vision (ECCV)}}, pages
  740--755. Springer, 2014.

\bibitem{MPJ13}
Hodosh Micah, Young Peter, and Hockenmaier Julia.
\newblock {Framing Image Description as a Ranking Task: Data, Models and
  Evaluation Metrics}.
\newblock {\em {Journal of Artificial Intelligence Research}}, 2013.

\bibitem{NDRSMMSC21}
Alex Nichol, Prafulla Dhariwal, Aditya Ramesh, Pranav Shyam, Pamela Mishkin,
  Bob McGrew, Ilya Sutskever, and Mark Chen.
\newblock {GLIDE: Towards Photorealistic Image Generation and Editing with
  Text-Guided Diffusion Models}.
\newblock {\em {CoRR abs/2112.10741}}, 2021.

\bibitem{RKHRGASAMCKS21}
Alec Radford, Jong~Wook Kim, Chris Hallacy, Aditya Ramesh, Gabriel Goh,
  Sandhini Agarwal, Girish Sastry, Amanda Askell, Pamela Mishkin, Jack Clark,
  Gretchen Krueger, and Ilya Sutskever.
\newblock {Learning Transferable Visual Models From Natural Language
  Supervision}.
\newblock In {\em {International Conference on Machine Learning (ICML)}}, pages
  8748--8763. PMLR, 2021.

\bibitem{RDNCC22}
Aditya Ramesh, Prafulla Dhariwal, Alex Nichol, Casey Chu, and Mark Chen.
\newblock {Hierarchical Text-Conditional Image Generation with CLIP Latents}.
\newblock {\em {CoRR abs/2204.06125}}, 2022.

\bibitem{RPGGVRCS21}
Aditya Ramesh, Mikhail Pavlov, Gabriel Goh, Scott Gray, Chelsea Voss, Alec
  Radford, Mark Chen, and Ilya Sutskever.
\newblock {Zero-Shot Text-to-Image Generation}.
\newblock In {\em {International Conference on Machine Learning (ICML)}}, pages
  8821--8831. JMLR, 2021.

\bibitem{RAYLSL16}
Scott~E. Reed, Zeynep Akata, Xinchen Yan, Lajanugen Logeswaran, Bernt Schiele,
  and Honglak Lee.
\newblock {Generative Adversarial Text to Image Synthesis}.
\newblock In {\em {International Conference on Machine Learning (ICML)}}, pages
  1060--1069. JMLR, 2016.

\bibitem{RG19}
Nils Reimers and Iryna Gurevych.
\newblock {Sentence-BERT: Sentence Embeddings using Siamese BERT-Networks}.
\newblock In {\em {Conference on Empirical Methods in Natural Language
  Processing and International Joint Conference on Natural Language Processing
  (EMNLP-IJCNLP)}}, pages 3980--3990. ACL, 2019.

\bibitem{RBLEO22}
Robin Rombach, Andreas Blattmann, Dominik Lorenz, Patrick Esser, and
  Bj{\"{o}}rn Ommer.
\newblock {High-Resolution Image Synthesis with Latent Diffusion Models}.
\newblock In {\em {IEEE Conference on Computer Vision and Pattern Recognition
  (CVPR)}}, pages 10684--10695. IEEE, 2022.

\bibitem{SCSLWDGAMLSHFN22}
Chitwan Saharia, William Chan, Saurabh Saxena, Lala Li, Jay Whang, Emily
  Denton, Seyed Kamyar~Seyed Ghasemipour, Burcu~Karagol Ayan, S.~Sara Mahdavi,
  Rapha~Gontijo Lopes, Tim Salimans, Jonathan Ho, David~J. Fleet, and Mohammad
  Norouzi.
\newblock {Photorealistic Text-to-Image Diffusion Models with Deep Language
  Understanding}.
\newblock {\em {CoRR abs/2205.11487}}, 2022.

\bibitem{SDTLH22}
Shibani Santurkar, Yann Dubois, Rohan Taori, Percy Liang, and Tatsunori
  Hashimoto.
\newblock {Is a Caption Worth a Thousand Images? {A} Controlled Study for
  Representation Learning}.
\newblock {\em {CoRR abs/2207.07635}}, 2022.

\bibitem{SVBKMKCJK21}
Christoph Schuhmann, Richard Vencu, Romain Beaumont, Robert Kaczmarczyk,
  Clayton Mullis, Aarush Katta, Theo Coombes, Jenia Jitsev, and Aran
  Komatsuzaki.
\newblock {{LAION-400M:} Open Dataset of CLIP-Filtered 400 Million Image-Text
  Pairs}.
\newblock {\em {CoRR abs/2111.02114}}, 2021.

\bibitem{SWR20}
Douglas~M. Souza, Jonatas Wehrmann, and Duncan~D. Ruiz.
\newblock {Efficient Neural Architecture for Text-to-Image Synthesis}.
\newblock In {\em {International Joint Conference on Neural Networks (IJCNN)}},
  pages 1--8. IEEE, 2020.

\bibitem{VSPUJGKP17}
Ashish Vaswani, Noam Shazeer, Niki Parmar, Jakob Uszkoreit, Llion Jones,
  Aidan~N. Gomez, Lukasz Kaiser, and Illia Polosukhin.
\newblock {Attention is All you Need}.
\newblock In {\em {Annual Conference on Neural Information Processing Systems
  (NIPS)}}, pages 5998--6008. NIPS, 2017.

\bibitem{WWZOE20}
Sheng{-}Yu Wang, Oliver Wang, Richard Zhang, Andrew Owens, and Alexei~A. Efros.
\newblock {CNN-Generated Images Are Surprisingly Easy to Spot... for Now}.
\newblock In {\em {IEEE Conference on Computer Vision and Pattern Recognition
  (CVPR)}}, pages 8692--8701. IEEE, 2020.

\bibitem{WQWHC20}
Zixu Wang, Zhe Quan, Zhi{-}Jie Wang, Xinjian Hu, and Yangyang Chen.
\newblock {Text to Image Synthesis With Bidirectional Generative Adversarial
  Network}.
\newblock In {\em {International Conference on Multimedia and Expo (ICME)}},
  pages 1--6. IEEE, 2020.

\bibitem{YLHH14}
Peter Young, Alice Lai, Micah Hodosh, and Julia Hockenmaier.
\newblock {From image descriptions to visual denotations: New similarity
  metrics for semantic inference over event descriptions}.
\newblock {\em {Transactions of the Association for Computational
  Linguistics}}, 2014.

\bibitem{YDF19}
Ning Yu, Larry Davis, and Mario Fritz.
\newblock {Attributing Fake Images to GANs: Learning and Analyzing {GAN}
  Fingerprints}.
\newblock In {\em {IEEE International Conference on Computer Vision (ICCV)}},
  pages 7555--7565. IEEE, 2019.

\bibitem{ZKBLY21}
Han Zhang, Jing~Yu Koh, Jason Baldridge, Honglak Lee, and Yinfei Yang.
\newblock {Cross-Modal Contrastive Learning for Text-to-Image Generation}.
\newblock In {\em {IEEE Conference on Computer Vision and Pattern Recognition
  (CVPR)}}, pages 833--842. IEEE, 2021.

\bibitem{ZXL17}
Han Zhang, Tao Xu, and Hongsheng Li.
\newblock {StackGAN: Text to Photo-Realistic Image Synthesis with Stacked
  Generative Adversarial Networks}.
\newblock In {\em {IEEE International Conference on Computer Vision (ICCV)}},
  pages 5908--5916. IEEE, 2017.

\bibitem{ZKC192}
Xu~Zhang, Svebor Karaman, and Shih{-}Fu Chang.
\newblock {Detecting and Simulating Artifacts in {GAN} Fake Images}.
\newblock In {\em {IEEE International Workshop on Information Forensics and
  Security (WIFS)}}, pages 1--6. IEEE, 2019.

\end{thebibliography}
\end{small}
%-----------------------------------------------------

%%%%%%%%%%%%%%%%%%%%%%%%%%%%%%%%%%%%%%%%%%%%%%%%%%%%%%%%%%%%%%%%%%%%%%%%%%%%%%%%
\end{document}